\RequirePackage{fix-cm}
\documentclass[a4wide,numbers=noendperiod]{scrartcl}
\usepackage{graphicx}
\usepackage{natbib}

\usepackage{siunitx}

\usepackage{booktabs}
\usepackage{tabularx}
\newcolumntype{C}{>{\centering\arraybackslash}X}
\newcolumntype{R}{>{\raggedleft\arraybackslash}X}
\newcolumntype{L}{>{\raggedright\arraybackslash}X}

\usepackage{float}  

\usepackage[breaklinks,hidelinks]{hyperref}

\usepackage{algorithm}
\usepackage{algorithmic}

\usepackage[usenames,dvipsnames,table]{xcolor}

\usepackage{subfig}

\usepackage{amsmath}
\usepackage{amssymb}
\usepackage{amsthm}

\newtheorem{definition}{Definition}

\begin{document}

\title{Experimental Analysis of a Generalized Stratified Sampling Algorithm for Hypercubes
}


\author{Simon Wessing\\Chair of Algorithm Engineering\\Computer Science Department\\Technische Universit\"at Dortmund, Germany\\\texttt{simon.wessing@tu-dortmund.de}}



\date{}

\maketitle

\begin{abstract}
Stratified sampling is a fast and simple method to generate point sets with uniform distribution in hypercubes. However, for the most common paraxial stratfication it has the prominent drawback that the number of sampled points in $n$ dimensions has to be an $n$-th power of an integer number. 
This exponential growth makes its application unattractive or even infeasible in high dimensions.
We present a stratification procedure that eliminates this problem by a recursive binary partitioning of the hypercube. 
The algorithm runs in linear time and tries to minimize the hyperboxes' deviation from the cubic shape. 
We analyze the properties of the algorithm using discrepancy and covering radius, and directly in practical applications, comparing it to quasirandom and other sampling methods.
We also discuss a potential combination with Latin hypercube sampling, which positively affects discrepancy.
\end{abstract}

\section{Introduction}

Stratified sampling is a classic strategy for generating point sets with uniform distribution in hypercubes. 
The general idea of the algorithm is to partition the space into disjunct strata and then to sample one point per stratum. 
With an appropriate choice of strata, e.\,g., as cells of a paraxial grid, a linear runtime can be achieved for this procedure.
Thus, it is among the methods that are especially suited for large sample sizes.
It has potential applications, e.\,g., in rendering of computer graphics~\cite[pp.~302-315]{Pharr2004} or in motion-planning of robots~\cite[pp.~185-209]{LaValle2006}.

It can be shown that stratified sampling yields Monte Carlo estimators whose variance is not larger than that of estimators based on random uniform sampling~\citep{Cheng1989}.
In the seminal paper of \cite{McKay1979}, also the related approach of Latin hypercube sampling (LHS) was introduced and obtained a more favorable evaluation.
An important drawback of stratified sampling with a paraxial grid structure is that the number of points $N$ cannot be chosen freely.
If the resolution (the number of bins) of the grid is the same in each dimension, $N$ has to be an $n$-th power of this integer number.
Likewise, if the number of bins is allowed to vary across dimensions, we still have to find a factorization of $N$ into $n$ integers, and ideally we would need problem knowledge to decide which dimensions to resolve finer or coarser.
In this work, we will show a possible way for giving up the grid structure, and thus enabling an arbitrary number of points without extra requirements. 
Additionally, we show how sampling each stratum with a non-uniform distribution can improve the covering radius of the resulting point set.

\begin{figure*}[t]
\centering

\subfloat[Latin hypercube sampling]{\includegraphics[width=0.32\textwidth]{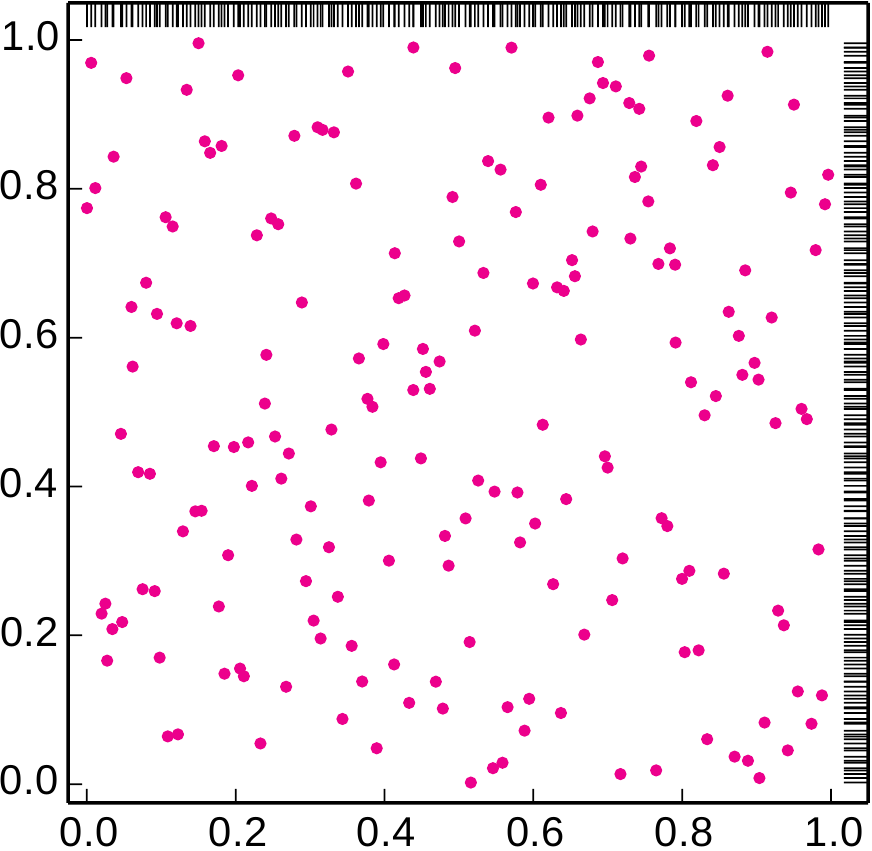}\label{fig:point_set_examples_lhs}}
\hfill\subfloat[Halton sequence]{\includegraphics[width=0.32\textwidth]{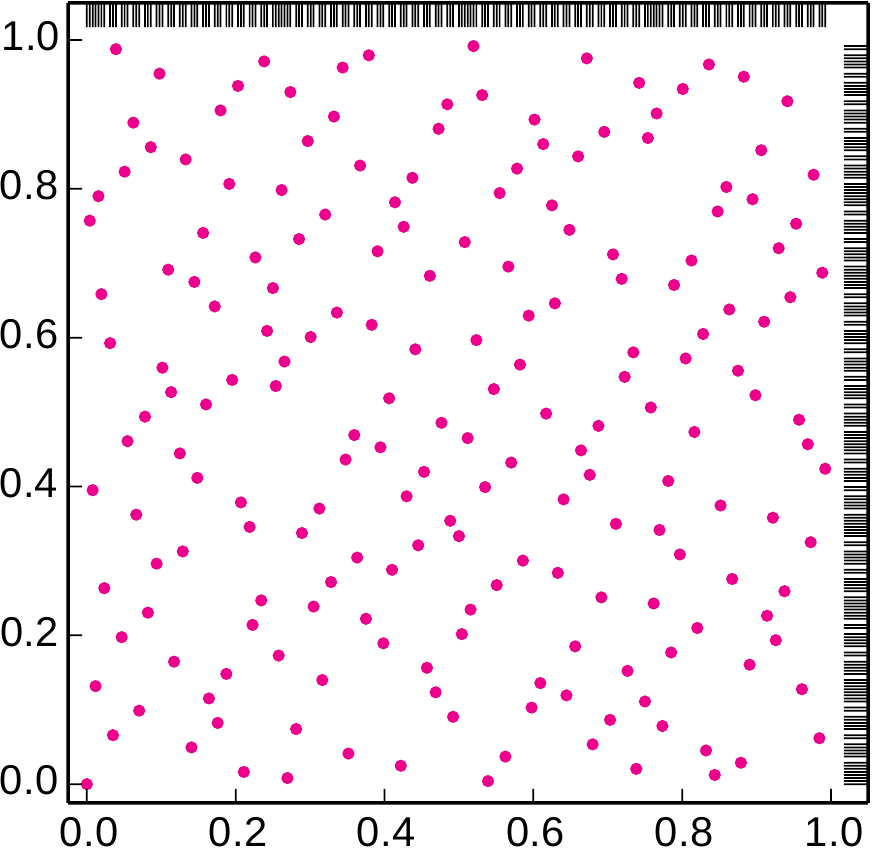}\label{fig:point_set_examples_halton}}
\hfill\subfloat[Stratified sampling]{\includegraphics[width=0.32\textwidth]{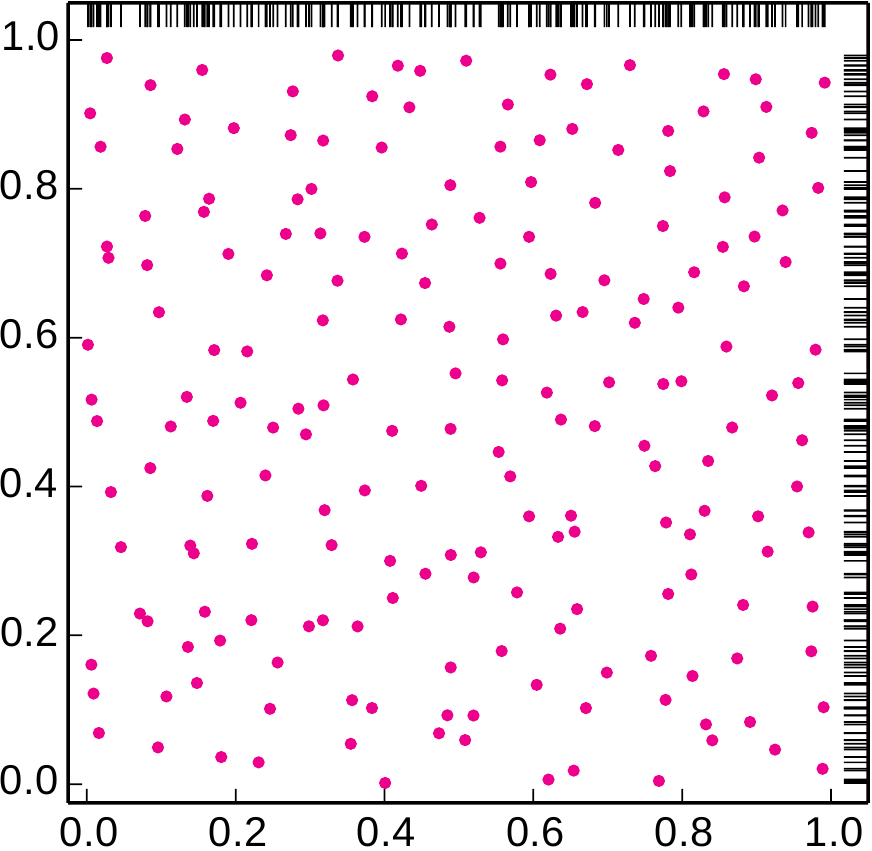}\label{fig:point_set_examples_stratified_general}}
%
%
%
\caption{Comparison of Latin hypercube sampling, the Halton sequence, and stratified sampling with $N = 196$ points. 
One-dimensional projections are indicated at the upper and right sides, respectively.}
\label{fig:comparison_with_other}
\end{figure*}

For simplicity we will assume that our region of interest is $X = [0,1]^n \subset \mathbb{R}^n$, and we want to uniformly sample $N$ points in it. 
However, note that most methods discussed here also work on arbitrary hypercubes.
We also put the main focus of our investigations on algorithms with a linear runtime in the number of points, as this is often the only affordable class for large sample sizes.
To begin with a graphic example, Fig.~\ref{fig:comparison_with_other} compares a stratified sampling in the sense of~\cite{McKay1979} with Latin hypercube sampling and the Halton sequence~\citep{Halton1964}. 
The distributions of the Halton sequence and stratified sampling are more uniform than LHS, but the Halton sequence and LHS have more uniform one-dimensional projections than the stratified points, i.\,e., when we divide any dimension equally into $N$ bins, every bin contains exactly one point. 
LHS always possesses this property by construction, the Halton sequence only for special values of~$N$.
Both the uniform $n$-dimensional and one-dimensional distributions are desirable properties, and by investigating different variants of stratified sampling we indirectly also gain insights of how they affect experimental results in numerical integration and worst-case optimization.

The outline of the paper is as follows.
We first give some pointers to related research in Sect.~\ref{sec:related_research}, before we describe the proposed algorithm in Sect.~\ref{sec:algo_description}. 
In Sect.~\ref{sec:summary_characteristics} we explain why the two summary characteristics discrepancy and covering radius are relevant for the assessment of uniform point sets, and how the stratification can be used to efficiently calculate an upper bound for the covering radius.
In Sect.~\ref{sec:experiments} we experimentally investigate an implementation of the proposed methodology, before we draw conclusions in Sect.~\ref{sec:conclusion}.

\section{Related Work on Stratified Sampling}
\label{sec:related_research}

Several existing approaches generalize stratified sampling in some way or another, to overcome its restriction regarding the number of points and to improve performance. 
For example, \cite{Cheng1989} propose to sample a one- or two-dimensional space of control variables instead of the original space.
However, suitable control variables are often not available.
Recursive stratified sampling~\cite[pp.~316--328]{Press1992} is another approach. 
It is an adaptive Monte Carlo method, recursively dividing the space into subregions by splitting along one dimension at a time. 
The location of the division is chosen to minimize the variance in the new subregions, which is estimated from samples drawn previously in the subregion.
More points are sampled in regions of higher variance, so the resulting distribution is non-uniform.
Ideas from recursive stratified sampling have later been adopted in much more sophisticated algorithms for rendering of computer graphics~\citep{Agarwal2003,Hachisuka2008}.

\cite{Shields2016} present a whole taxonomy, with conventional stratified sampling and LHS as extreme cases of a class of stratified sampling algorithms.
Between the two extremes, they locate partially stratified sample (PSS) designs, which are simply assembled from lower-dimen\-sio\-nal designs.
This approach yields more degrees of freedom for the number of points $N$, similarly as allowing a varying number of points per dimension does~\citep{Chiu1994}.
To obtain an arbitrary $N$, \cite{Kensler2013} proposes (for the 2-D special case) to choose $mn \geq N$, sample $mn$ points in the unit hypercube, and then stretch the design in the dimension that was sampled more densely, so that only $N$ points lie in the unit hypercube.
In the following, we will propose yet another approach for obtaining an arbitrary $N$.

\section{Algorithm Description}
\label{sec:algo_description}

Our algorithm is inspired by the part-and-select algorithm (PSA) of \cite{Salomon2013}, which is geared to the efficient selection of a uniform subset of a given point set.
PSA itself is closely related to several algorithms for vector quantization, as the median-cut algorithm by \cite{Heckbert1982}, the mean-split algorithm by \cite{Wu1985}, and the method of \cite{Wan1988}. 
These approaches are all recursive partitioning algorithms using hyperboxes to describe the clusters. 
The main difference between them is the criterion determining where a cluster is split in two. 
In contrast to the other algorithms, PSA does not aim to minimize the quantization error, but simply to obtain a uniform subset of the original data. 
It was developed originally for subset selection in multiobjective optimization, but there are no special assumptions that prevent a universal application. 

\begin{algorithm}[t]
\begin{algorithmic}[1]
\REQUIRE number of strata $N$, number of variables $n$, lower bounds $\vec{l} = (\ell_1, \dots, \ell_n)^\top$, upper bounds $\vec{u} = (u_1, \dots, u_n)^\top$
\ENSURE partition of the space $A$
\STATE $S_1 \gets (\vec{l}, \vec{u})$ \COMMENT{initialize first stratum with hypercube}
\STATE $N_{1} \gets N$ \COMMENT{assign all points to it}
\STATE $A = \{S_1\}$ \COMMENT{init with currently existing strata}
\WHILE[if stratum with $>1$ points exists\!]{$\exists S_j \in A: N_{j} > 1$}
\STATE $N_a \gets \lfloor N_{j} / 2\rfloor$ \COMMENT{divide as evenly as possible}
\IF{$(N_j \geq 6) \wedge (N_a \bmod 2 \neq 0) \wedge (N_j \bmod 2 = 0)$}
\STATE $N_a \gets N_a - 1$ \COMMENT{avoid odd numbers (optional)}
\ENDIF
\STATE $N_b \gets N_j - N_a$
\STATE $s \gets \arg\max\{u_{j,i} - \ell_{j,i} \mid i = 1, \dots, n\}$ \COMMENT{identify longest side\!\!} \label{alg:longest_side}
\STATE $p_s \gets \ell_{j,s} + (u_{j,s} - \ell_{j,s}) \cdot N_a / N_{j}$ \COMMENT{calculate split position}
\STATE $\vec{u}_{a} \gets (u_{j,1}, \dots, u_{j,s-1}, p_s, u_{j,s+1}, \dots, u_{j,n})^\top$
\STATE $\vec{l}_{b} \gets (\ell_{j,1}, \dots, \ell_{j,s-1}, p_s, \ell_{j,s+1}, \dots, \ell_{j,n})^\top$
\STATE $S_a \gets (\vec{l}_{j}, \vec{u}_{a})$ \COMMENT{stratum with updated upper bounds}
\STATE $S_b \gets (\vec{l}_{b}, \vec{u}_{j})$ \COMMENT{stratum with updated lower bounds}
\STATE $A \gets A \setminus \{S_j\}$ \COMMENT{remove current stratum}
\STATE $A \gets A \cup \{S_a, S_b\}$ \COMMENT{add the two smaller ones~}
\ENDWHILE
\RETURN $A$
\end{algorithmic}
\caption{Stratification of a hypercube with arbitrary number of points}
\label{alg:strat_sampling}
\end{algorithm}

In~\cite[pp.~63--68]{Wessing2015}, it was noticed that subset selection algorithms can be easily repurposed for sampling, by generating random uniform points and selecting from them.
However, the resulting distributions may exhibit subtle deviations from uniformity, especially in the boundary region of the hypercube, and the resulting runtime is super-linear. 
Thus, the stratified sampling as presented in Alg.~\ref{alg:strat_sampling} was developed based on a central aspect of  PSA and its ancestors, namely the splitting of a hyperbox in half at its longest side. 
We will call it generalized stratified sampling (GSS) here, because it can also be combined with PSS and thus yields an even larger design space of stratified point sets.
Contrarily to PSA, GSS does not have to keep track to which hyperbox each point belongs, as the points are not yet existing.
Instead, each stratum is assigned a number of points to be later sampled in it.
During the partitioning, the algorithm maintains the invariant that the relative volume of the strata is proportional to the number of points assigned to them. 
Thus, in case a stratum contains an odd number of points, the side lengths of the resulting strata after the split cannot be exactly half of the previous one, but have to correspond to the integer numbers of points assigned to each stratum.
The partitioning is continued until each stratum is assigned exactly one point.
At this stage, each stratum has a volume of $1/N$ of the initial hypercube.
Finally, the strata are returned and a random uniform point can be sampled from each one. 
Given that the used strata are saved, the $N$ sampled points can also be later augmented in linear time by $kN$ new points, while maintaining the uniform distribution.

\begin{figure*}[t]
\centering

\subfloat[Conventional (minimal ratio of side lengths: 1)]{\includegraphics[width=0.32\textwidth]{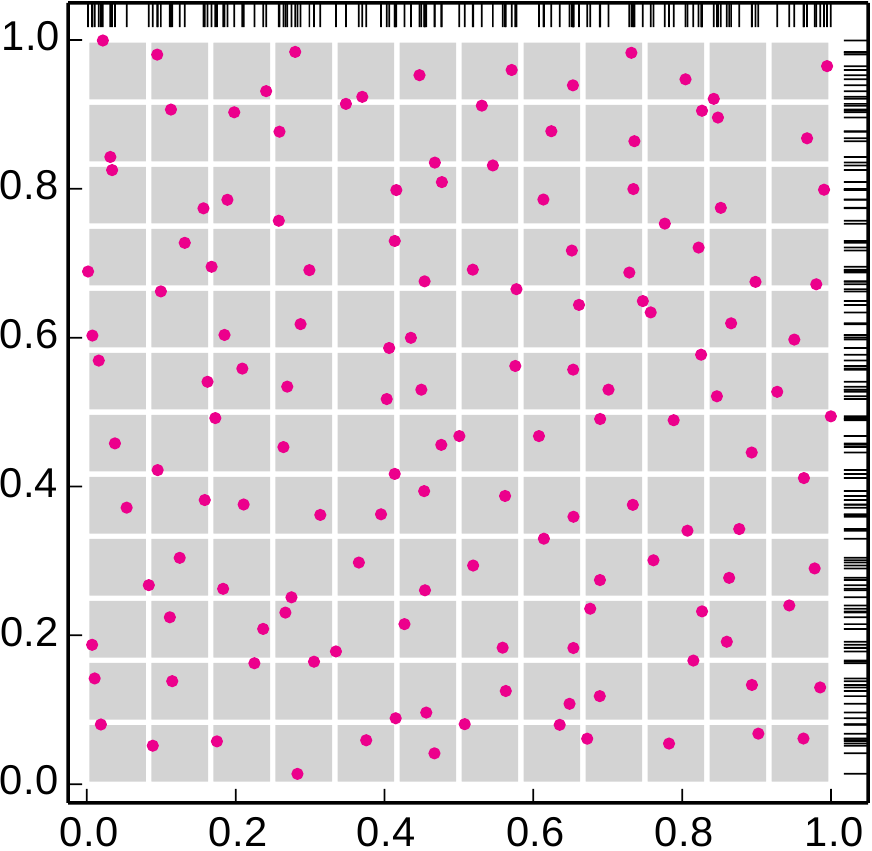}\label{fig:point_set_examples_conventional_strat}}
\hfill\subfloat[Alg.~\ref{alg:strat_sampling} without lines 6--7 (minimal ratio of side lengths: $1/2.\bar{7}$)]{\includegraphics[width=0.32\textwidth]{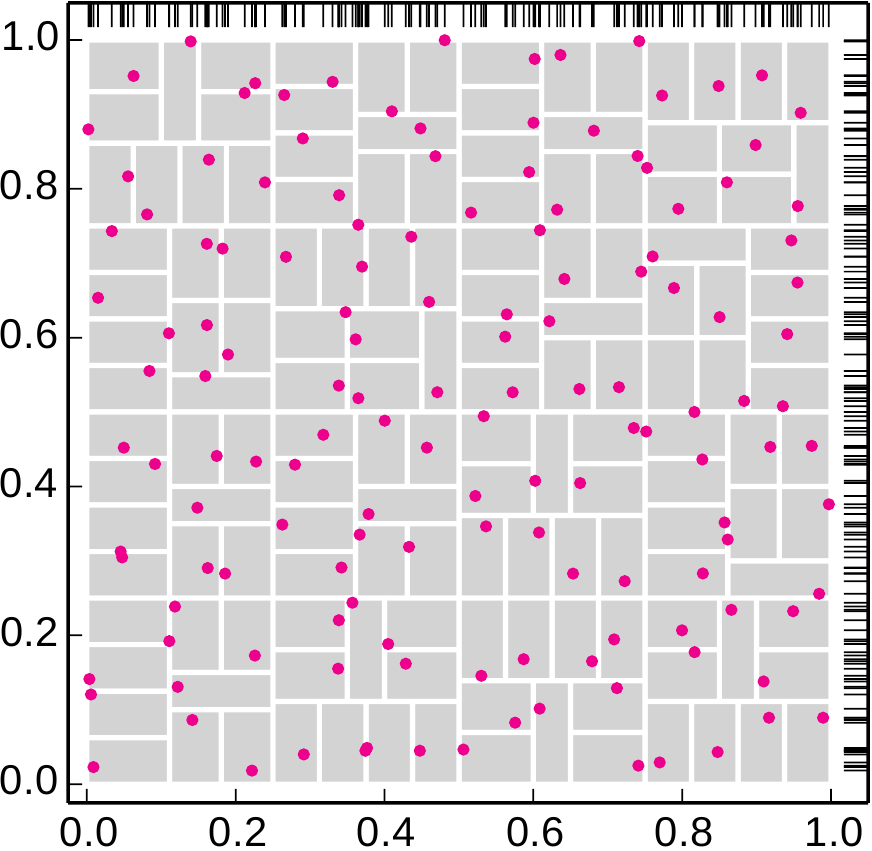}\label{fig:point_set_examples_our_strat_wo_avoid_odd}}
\hfill\subfloat[Alg.~\ref{alg:strat_sampling} with lines 6--7 (minimal ratio of side lengths: $1/1.\bar{7}$)]{\includegraphics[width=0.32\textwidth]{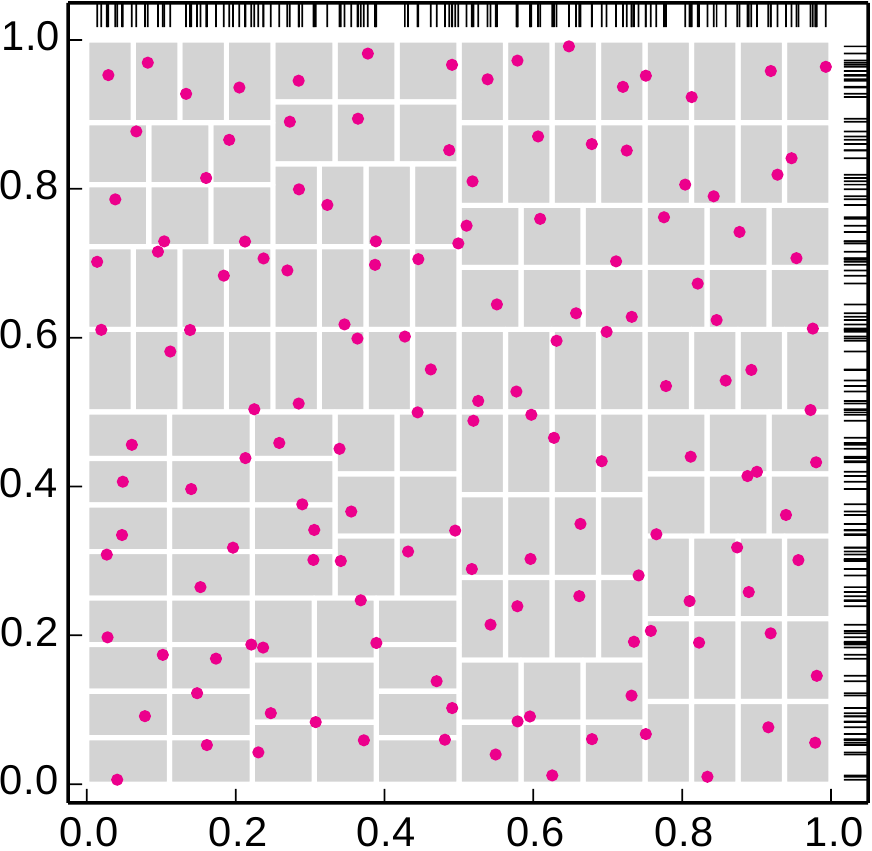}\label{fig:point_set_examples_our_strat_w_avoid_odd}}

\caption{Comparison of conventional stratified sampling and GSS with $N = 144$ points. 
The stratification is illustrated in gray.}
\label{fig:comparison_with_conventional}
\end{figure*}

Started on a hypercube, the algorithm also maintains the invariant that the ratio between the shortest and longest side of each stratum does not fall below $1/3$.
This is caused by the fact that the most uneven split can appear with three points left. 
For all splits involving more points, the split ratio is closer to $1/2$, and because we always split the longest side, a lower ratio cannot have appeared earlier either.
In many cases, it will be advisable to avoid splits resulting in an odd number of points in both new strata, to ultimately avoid the extreme case of a $1/3$ ratio (see Fig.~\ref{fig:comparison_with_conventional}). 
For example, six points should be split into four and two points, and not three and three, to avoid the 1/3 ratio in the next step.
Thus, we include lines 6--7 in the algorithm, to detect this situation and move one point to the other new stratum, if we have at least $6$ points left for the current stratum.

The linear runtime is achieved by proper bookkeeping of the final and the unfinished strata. 
This can either be done by using a double-ended queue and appending final and unfinished strata to different ends, or (preferably) by using two separate data structures.
In the implementation, care should be taken to randomize certain decisions, i.\,e., $N_a$ and~$N_b$ should be swapped randomly, and in line~\ref{alg:longest_side}, ties should be broken randomly as well.
An implementation of this algorithm is provided in the Python package diversipy \citep{diversipy2018}.

At values of $N = 2^{ni}$, $i \in \mathbb{N}$, the result of the algorithm is identical to conventional stratified sampling. 
To obtain the conventional stratification also for other $n$-th powers of arbitrary bases $x \in \mathbb{N}$, the usual algorithm has to be used instead.
Figure~\ref{fig:comparison_with_conventional} shows example outputs of the two algorithms for $N = 12^2$. 
In a real-world application, Alg.~\ref{alg:strat_sampling} would of course be rather employed for values of $N$ where the conventional algorithm is not available.

\subsection{Sampling with the Bates Distribution}

\begin{figure*}[t]
\centering

\subfloat[$N = 500$, $b = 1$]{\includegraphics[width=0.32\textwidth]{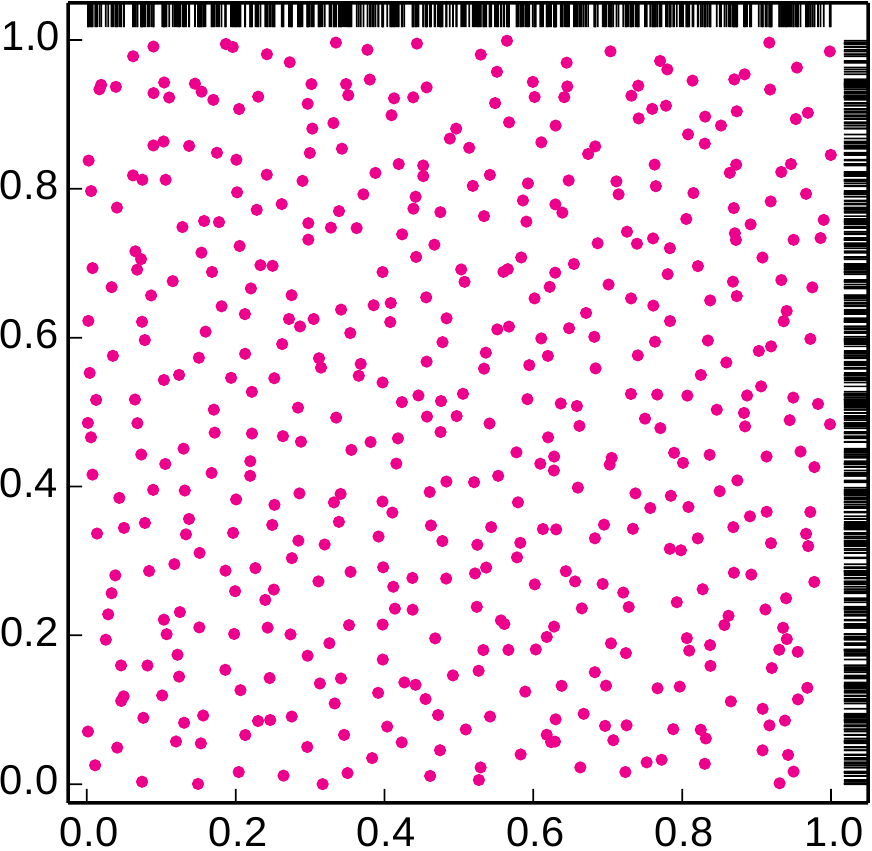}\label{fig:point_set_examples_stratified_500_1}}
\hfill\subfloat[$N = 500$, $b = 2$]{\includegraphics[width=0.32\textwidth]{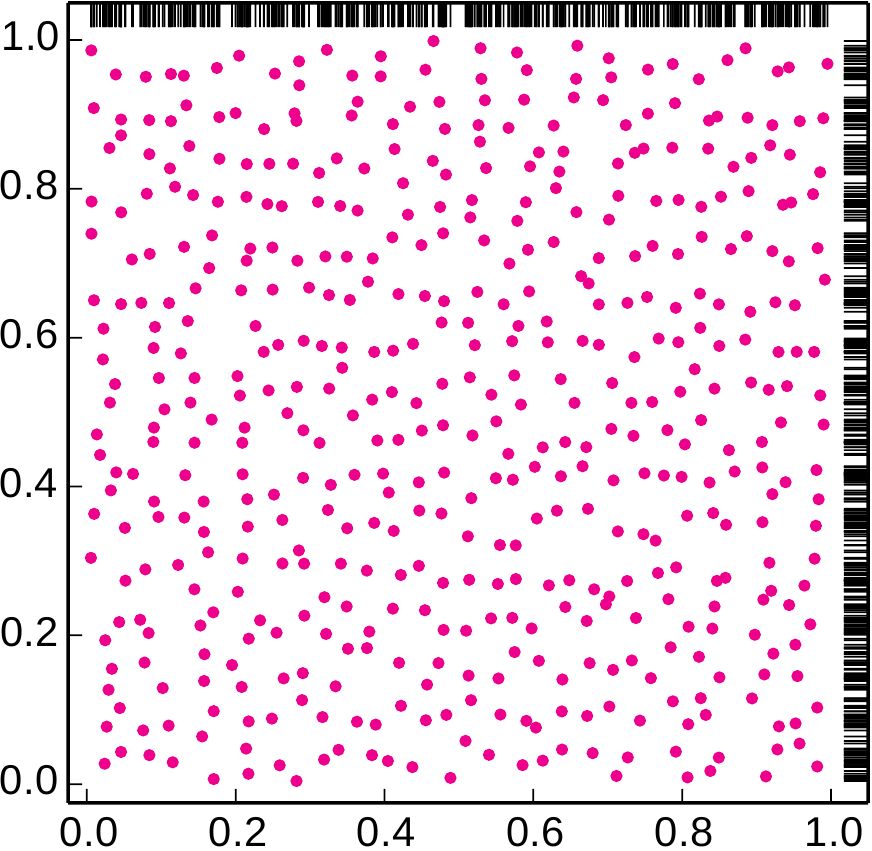}\label{fig:point_set_examples_stratified_500_2}}
\hfill\subfloat[$N = 500$, $b = 8$]{\includegraphics[width=0.32\textwidth]{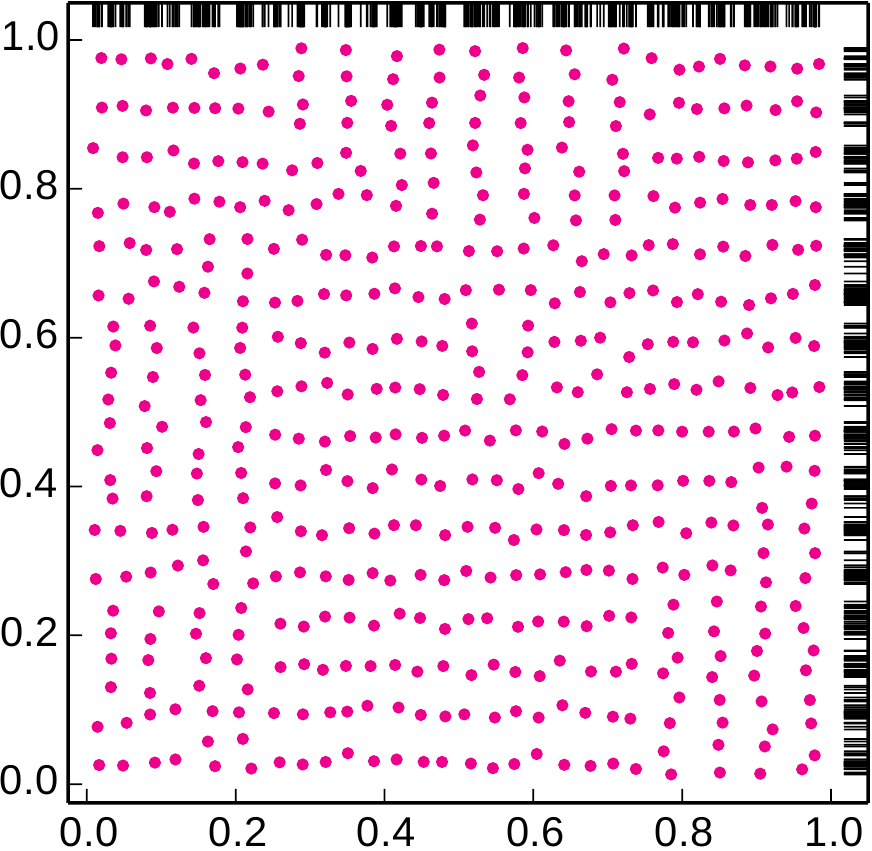}\label{fig:point_set_examples_stratified_500_8}}

\subfloat[$N = 500$, $b = \infty$]{\includegraphics[width=0.32\textwidth]{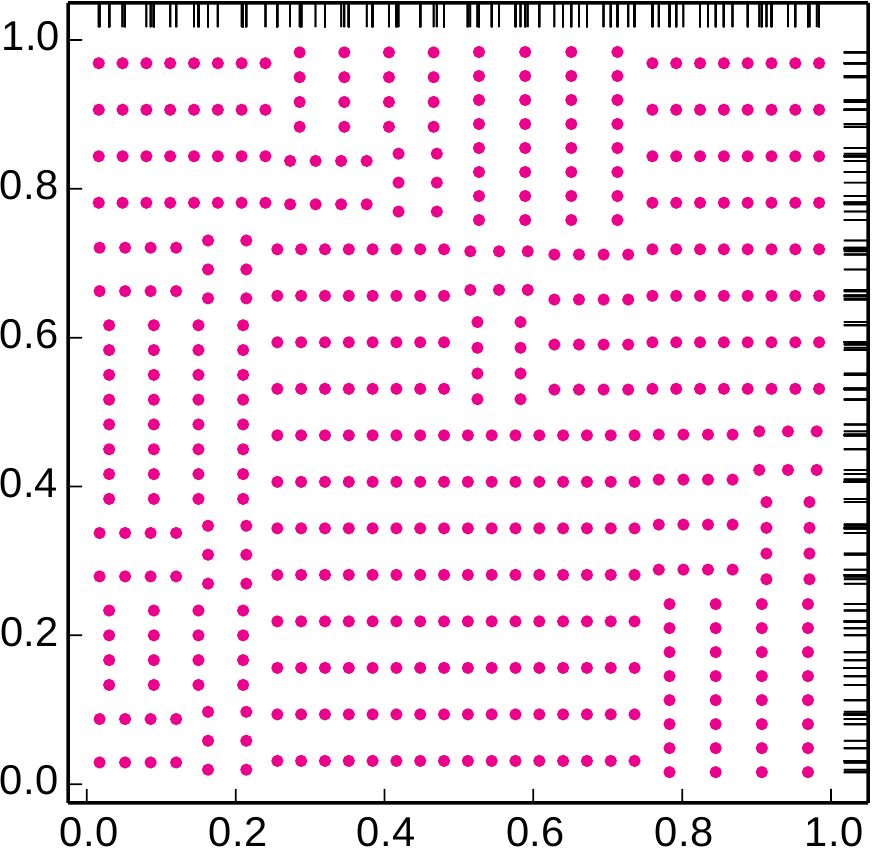}\label{fig:point_set_examples_stratified_500_inf}}
\hfill\subfloat[$N = 700$, $b = \infty$]{\includegraphics[width=0.32\textwidth]{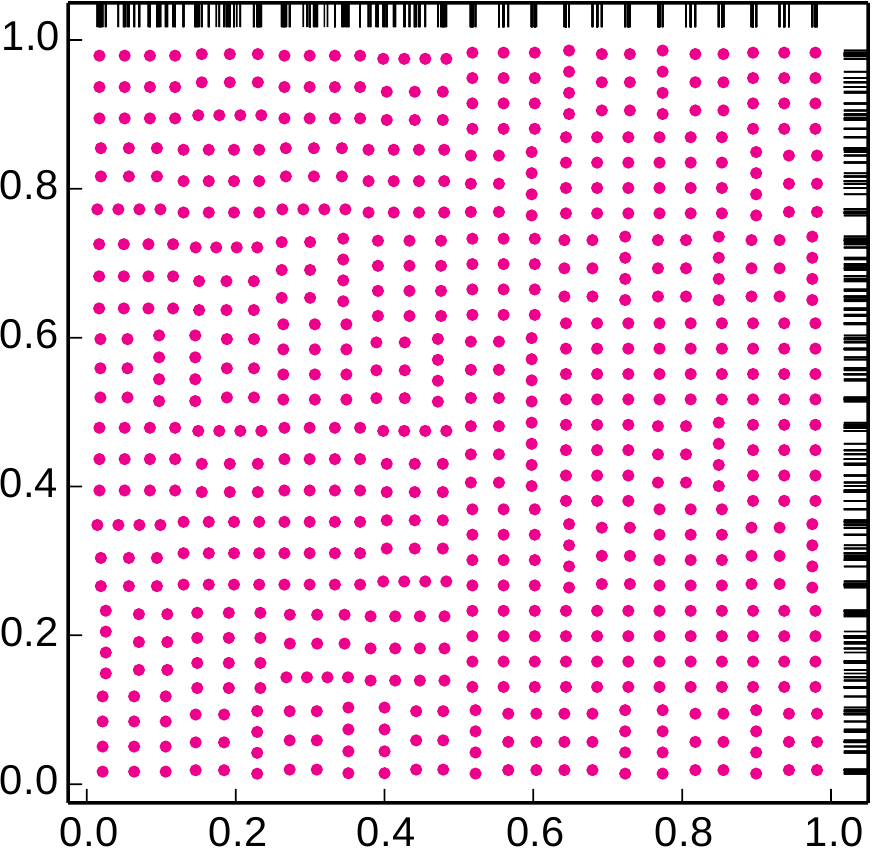}\label{fig:point_set_examples_stratified_700_inf}}
\hfill\subfloat[$N = 1000$, $b = \infty$]{\includegraphics[width=0.32\textwidth]{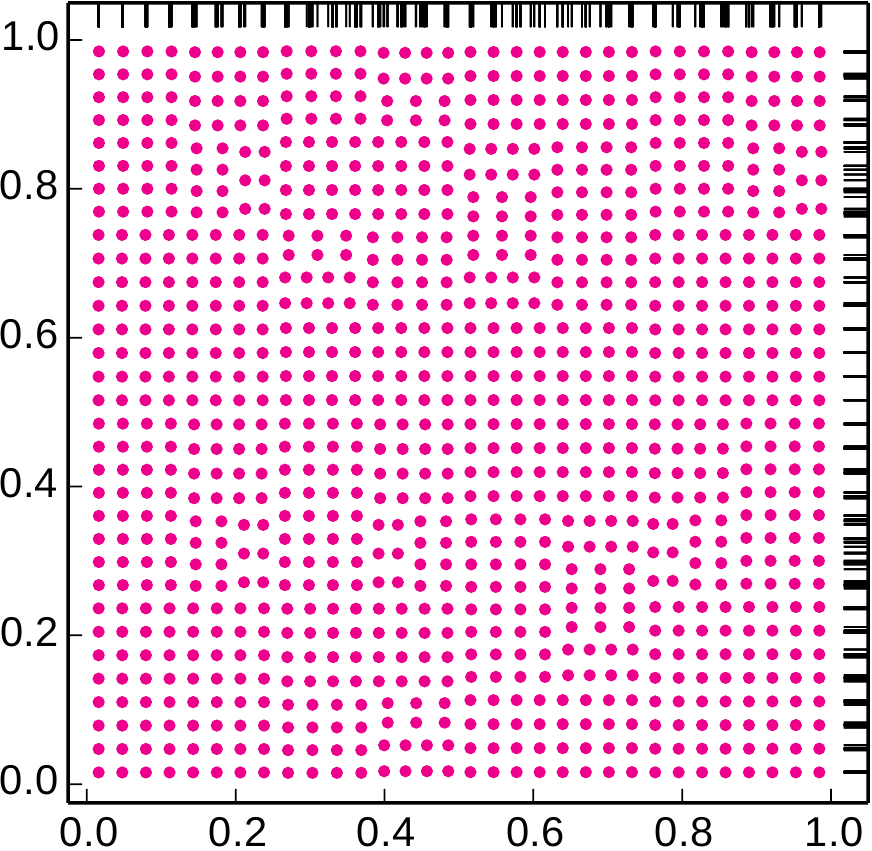}\label{fig:point_set_examples_stratified_1000_inf}}

%
%

\caption{Examples of GSS with different settings of $N$ and $b$. 
One-dimensional projections are indicated at the upper and right sides, respectively.}
\label{fig:bates_examples}
\end{figure*}

Alternatively to random uniform sampling, we could use the Bates distribution~\cite[p.~297]{Johnson1995} to obtain a larger expected separation distance between the points. 
Random numbers of this distribution are obtained by taking the mean of $b \in \mathbb{N}$ independent uniformly distributed random numbers. 
Thus, we can obtain a point $\vec{x} = (x_1, \dots, x_n)^\top$ by drawing
\[
x_i = \frac{1}{b} \sum_{j = 1}^b \mathcal{U}(\ell_i, u_i)\,,
\]
with $\vec{l} = (\ell_1, \dots, \ell_n)^\top$ being the lower and $\vec{u} = (u_1, \dots, u_n)^\top$ the upper bounds of a stratum.
$\mathcal{U}(\ell_i, u_i)$ is the uniform distribution, so it is guaranteed by construction that the point $\vec{x}$ does not exceed these boundaries. 
For $b = 1$, we retain random uniform sampling, while $b = \infty$ can be interpreted as deterministically taking the centroid of the stratum. 
This choice leads us to a sample identical to the Sukharev grid \citep{Sukharev1971} in the case of conventional stratified sampling.
Figure~\ref{fig:bates_examples} shows GSS examples with slightly larger point sets than in Fig.~\ref{fig:comparison_with_conventional} and different $b$ values. 
It is obvious that with increasing $b$, the distribution becomes more grid-like, which reduces the uniformity of the low-dimensional projections.


\subsection{Combination with Latinization}

For $b = 1$, the one-dimensional projections for stratified sampling are randomly uniform, but as shown in Fig.~\ref{fig:comparison_with_other}, a better distribution is possible.
For example, \cite{Saka2007} propose a procedure for converting arbitrary point sets into Latin hypercube (LH) designs. 
Quasirandom sequences possess good distributions in both spaces anyway~\citep{Kollig2002}, but may exhibit certain patterns that could be undesirable in some applications~\citep{Kensler2013}.
Thus, several authors already investigated combinations of stratified sampling and latin hypercube sampling.
\cite{Chiu1994} introduced the idea for 2-D spaces under the name \emph{multi-jittered sampling}.
\cite{Kensler2013} developed an online variant of the algorithm and improved its two-dimensional distribution by a slight derandomization.
Naturally, also PSS can be combined with latinization without effort~\citep{Shields2016}.

Note that due to the perfect alignment of strata in conventional stratified sampling, all approaches in the paragraph above run in linear time.
Unfortunately, adding latinization to GSS seems more difficult.
First of all, we observe that the problem can be divided into $n$ one-dimensional problems, because the Latin hypercube conditions apply to each dimension independently.
Each one-dimensional problem can then be formulated as a bipartite graph matching problem as follows: 
Let $B$ the bins of the LHS in one dimension and~$S$ the strata of the partitioning. 
Then construct a graph $G = (V, E)$ with vertices $V = B \cup S$, and edges $E$ representing intersecting bins and strata. 
A perfect matching of this graph represents an assignment of points to strata fulfilling the LH property.
Such a matching can for example be computed with the algorithm of~\cite{Hopcroft1973}, which has runtime $O(|E|\sqrt{|V|})$.
Doing this for every dimension and drawing the actual coordinates in the intersection of the assigned bin and stratum, we finally obtain our point set.
A similar approach appears in~\citep{Ahmed2016}, where a low-discrepancy point set is reorganized to add a blue noise property.

Unfortunately, with $|E| = O(N\sqrt{N})$ the runtime of this algorithm is $O(nN^2)$.
However, a valid assignment can often be obtained heuristically by sorting the strata by their center of gravity (COG), and assigning bins sequentially in this order. 
Our experiments indicate that the number of LH violations in this case is usually a single digit number, and they seem to become rarer in higher dimensions.
Using bucket sort, the sorting may be even accomplished in linear time if the COG distribution is uniform enough. 
Another (potential) benefit of this method is that some correlation between point coordinates and the strata's COG values is introduced, similarly to the method of~\cite{Kensler2013}. 
We will call this variant \emph{approximately latinized} GSS (ALGSS) in the following.
If a strict adherence to the LH property is desired, the heuristic solution could be used as an initialization for the graph matching algorithm, slashing its runtime considerably compared to a problem-agnostic greedy initialization.
However, as the results of the former approach are almost identical to those of ALGSS, we will consider a variant with randomized greedy initialization as LGSS in the experimental analysis in Sect.~\ref{sec:experiments}.

\section{Summary Characteristics}
\label{sec:summary_characteristics}

\subsection{Covering Radius}
\label{sec:covering_radius}

The \emph{covering radius} is an important measure for global optimization, because a worst-case error bound for the approximation of the global optimum can be given based on it~\cite[p.~149]{Niederreiter1992}. 
To keep this worst-case bound low, the covering radius should be minimized.

\begin{definition}[Covering radius]
\label{def:covering_radius}
The \emph{covering radius} of a point set $P = \{\vec{x}_1, \dots, \vec{x}_N\} \subset X$ is defined by
\begin{equation*}
d_\mathrm{cr}(P, X) = \sup_{\vec{x} \in X} \big\{\min_{1 \leq i \leq N} \{\|\vec{x} - \vec{x}_i\|_2\}\big\}\;. 
\end{equation*}
\end{definition}

\citet[p.~148]{Niederreiter1992} coined the term \emph{dispersion} for $d_\mathrm{cr}$, which sounds antithetic to the necessary minimization of this measure.
Thus, we will use the name \emph{covering radius}, which is used for example by \cite{Damelin2010}, because $d_\mathrm{cr}$ is the smallest radius for which closed balls around the points of $P$ completely cover~$X$. 
Note that in the area of computer experiments, Def.~\ref{def:covering_radius} is also known as minimax distance design criterion~\citep{Johnson1990}.

Unfortunately, it is quite difficult to calculate $d_\mathrm{cr}$ in general.
But \cite{Pronzato2012} give an algorithm for calculating $d_\mathrm{cr}(P, [0,1]^n)$ regarding Euclidean distance, based on Delaunay tessellation. 
It eludes us why the Delaunay tessellation is used in their description, as the Voronoi tessellation is a much more natural fit~\cite[p.~202]{LaValle2006}. 
The algorithm simply consists of mirroring the point set at all lower and upper boundaries of~$X$ and then computing the Voronoi tessellation of the multiplied point set.
The covering radius is then obtained by calculating the maximal distance of any Voronoi vertex to its nearest neighbor in $P$.
This algorithm has runtime $O((nN)^{\lfloor n/2 \rfloor})$, due to the Voronoi tessellation.


A lower bound can be obtained by using a Monte Carlo approach, because if $X$ is a finite point set with $|X| = M$, calculation of the measure becomes straightforward with runtime $O(nMN)$ \citep{Saka2007}. 
However, this lower bound is of limited use, because the covering radius is to be minimized.
In the following we will show that for stratified sampling, an upper bound can be computed in $O(nN)$, by calculating the distance of each point $\vec{x}$ to the furthest corner $\vec{y}^*$ of its stratum.
To obtain the runtime, it is important that the distance can be expressed as
\begin{equation*}
\label{eq:furthest_point_dist}
\|\vec{x} - \vec{y}^*\|_2 = \left(\sum_{i=1}^n \max\{x_i - \ell_i, u_i - x_i \}^2\right)^{1/2}\,,
\end{equation*}
with $\ell_i$ and $u_i$ denoting lower and upper bounds of the stratum%
\footnote{If this was not the case, there must be a dimension $i$ with $|x_i - y^*_i| \neq \max\{x_i - \ell_i, u_i - x_i \}$.
But if $|x_i - y^*_i|$ is larger, the point $\vec{y}^*$ would not be in the stratum, and if it is smaller there exists a point $\vec{y}'$ in the stratum that is further away.}. 
Note that $\|\vec{x} - \vec{y}^*\|_2$ is essentially the covering radius of a single point in its stratum.
To obtain the upper bound for $d_\mathrm{cr}$, it is sufficient to calculate 
\begin{equation}
\overline{d_\mathrm{cr}}(P, X) := \max\{d_\mathrm{cr}(\vec{x}_i, S_i) \mid i = 1, \dots, N\}\,,
\label{eq:cru}
\end{equation}
where $S_i$ are the strata with $\bigcup_{i=1}^N S_i = X$ and $\vec{x}_i \in S_i$ are the corresponding sample points. 
For the correctness of the upper bound the strata are not necessarily required to be disjunct. 
It suffices that the whole $X$ is covered and only one point is sampled per stratum.
In Sect.~\ref{sec:experiments}, we will show experimentally that this bound is often tight.

\subsection{Discrepancy}

In the area of quasi-Monte Carlo methods, a lot of theory has been developed regarding error bounds of estimated integrals, where the integrated function $f$ is treated as a black box. 
To achieve low error bounds, the used point sets must possess a low \emph{discrepancy}, as stated by the Koksma-Hlawka inequality \cite[p.~20]{Niederreiter1992}.
Intuitively, ``discrepancy can be viewed as a quantitative measure for the deviation from uniform distribution''~\citep[p.~13]{Niederreiter1992}, which should be as low as possible.
Several different variants can be defined by changing the aggregation of individual deviations or by considering differently shaped subsets of the region of interest.
We are considering the family $\mathcal{J}$ of subsets $J = [\ell_1, u_1) \times \dots [\ell_n, u_n)$ of the unit hypercube here, yielding the \emph{unanchored} discrepancy
\begin{equation*}
T_N = \left(\int_{(\vec{x},\vec{y}) \in X \times X,\, x_i < y_i} \! \left(\frac{N_J}{N} - \operatorname{vol}(J)\right)^2 \,\mathrm{d}\vec{x}\mathrm{d}\vec{y} \right)^{1/2}\,.
\end{equation*}
%
Here, $N_J$ is the number of points in subset $J$.
\cite{Morokoff1994} give the following explicit formula for $T_N$, which can be computed in $O(N^2n)$: 
\begin{alignat*}{2}
(T_N)^2 &= \frac{1}{N^2} \sum_{i=1}^N \sum_{j=1}^N \prod_{k=1}^n \big(1 - \max\{x_{i,k}, x_{j,k}\}\big) \cdot \min\{x_{i,k}, x_{j,k}\} \\
 &\phantom{=~} - \frac{2^{1-n}}{N} \sum_{i=1}^N \prod_{k=1}^n x_{i,k} (1 - x_{i,k}) + 12^{-n}\,. \nonumber
\end{alignat*}
\cite{Heinrich1996} offers an alternative algorithm with asymptotic runtime of only\linebreak$O(N(\log N)^n)$.
However, in practice an improvement is only obtained for $N > 2^{2n}$~\citep{Matousek1998}.
A useful property of discrepancy is the possibility to compute its expected value for a random uniform point set. 
For $(T_N)^2$ the formula is \citep{Morokoff1994} 
\begin{equation}
\operatorname{E}((T_N)^2) = \frac{6^{-n} (1 - 2^{-n})}{N} \;.
\label{eq:expected_value_disc}
\end{equation}
\cite{Hickernell1998} proposes several other variants of discrepancy that possess certain additional invariance properties. 
However, it seems that the corresponding expected values are unknown, so we will keep using $T_N$ instead.

\begin{figure*}[t!]
\centering

\subfloat[$n = 2$]{\includegraphics[width=0.47\textwidth]{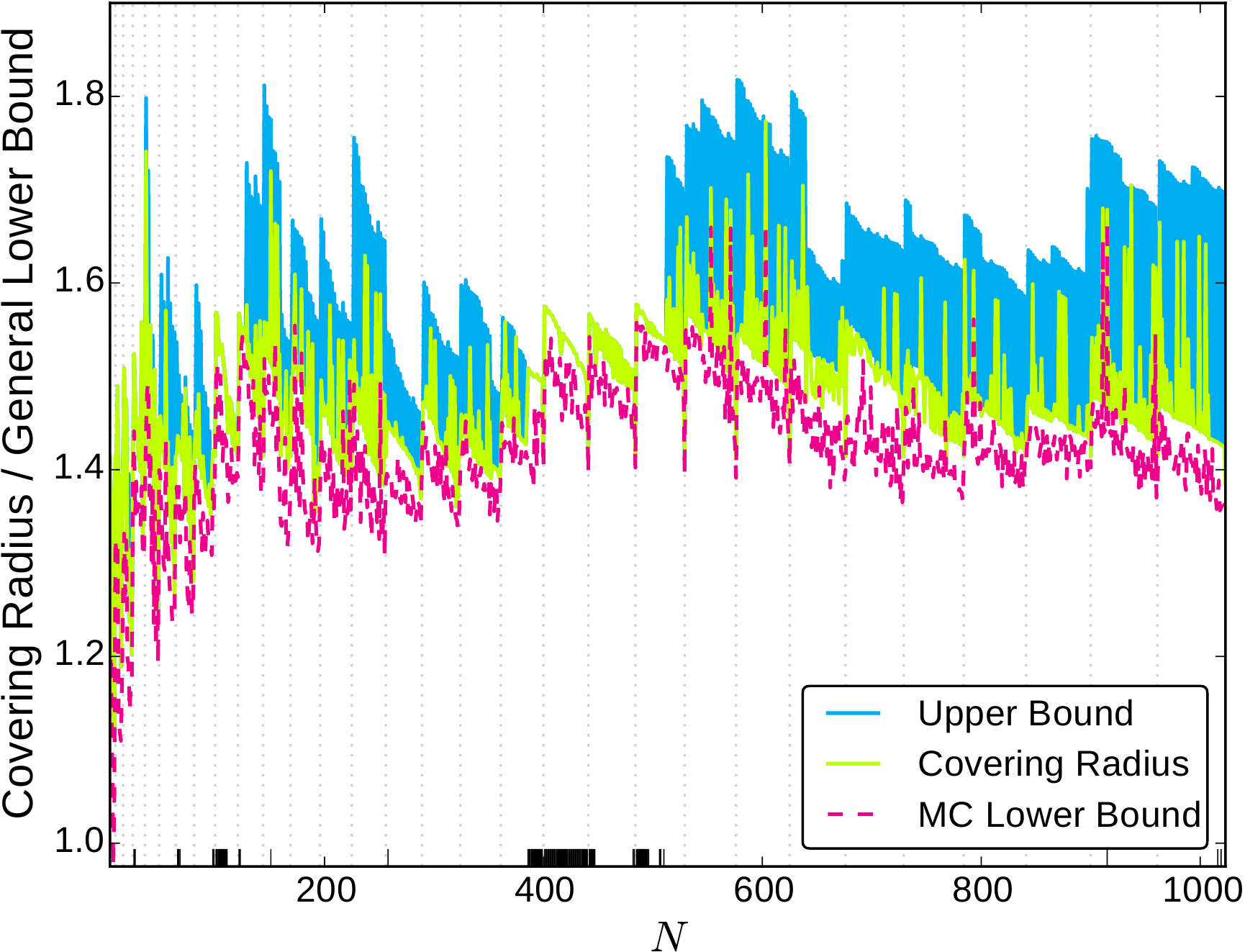}\label{fig:crbounds_vs_N10_dim2}}
\hfill\subfloat[$n = 3$]{\includegraphics[width=0.47\textwidth]{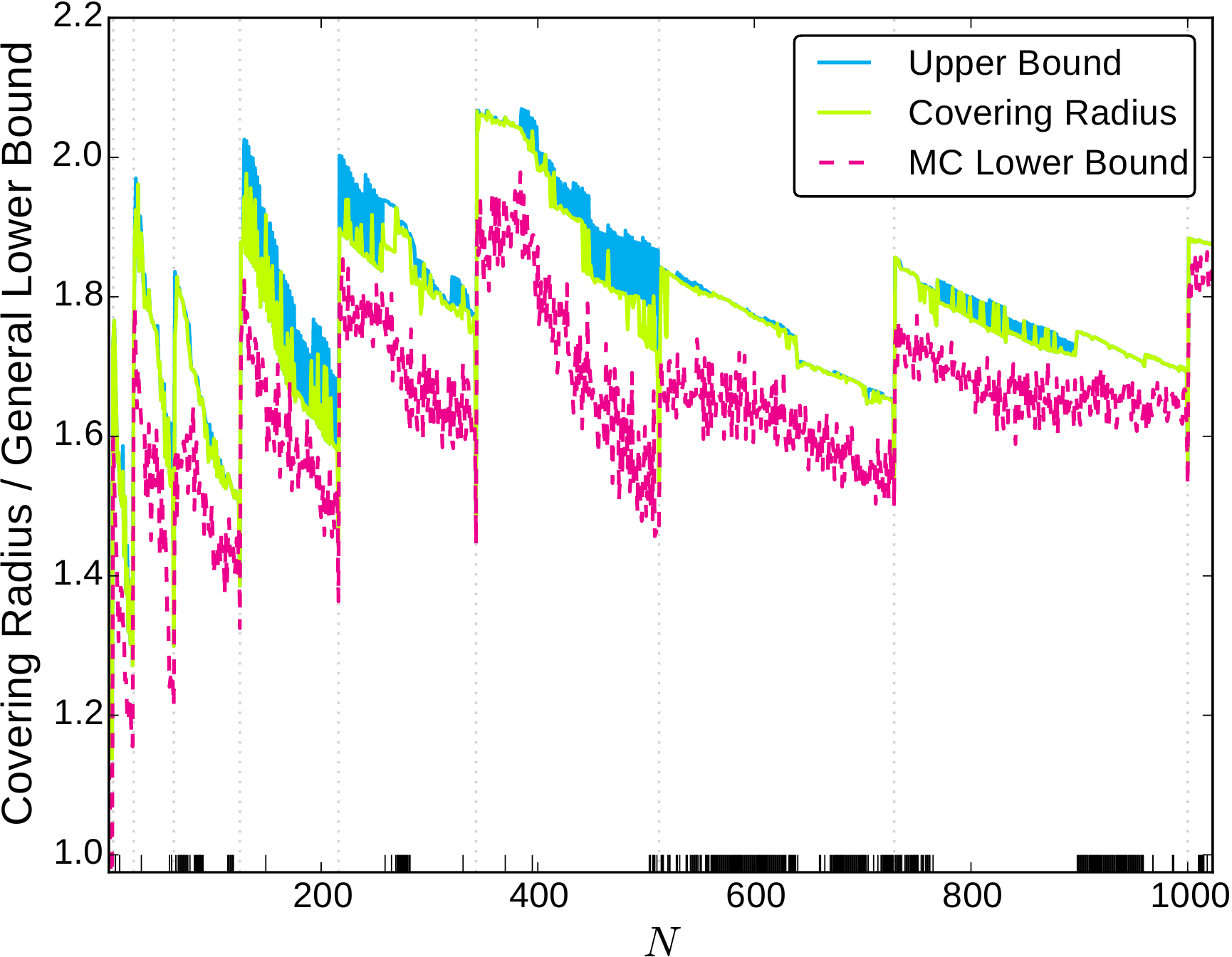}\label{fig:crbounds_vs_N10_dim3}}

\subfloat[$n = 5$]{\includegraphics[width=0.47\textwidth]{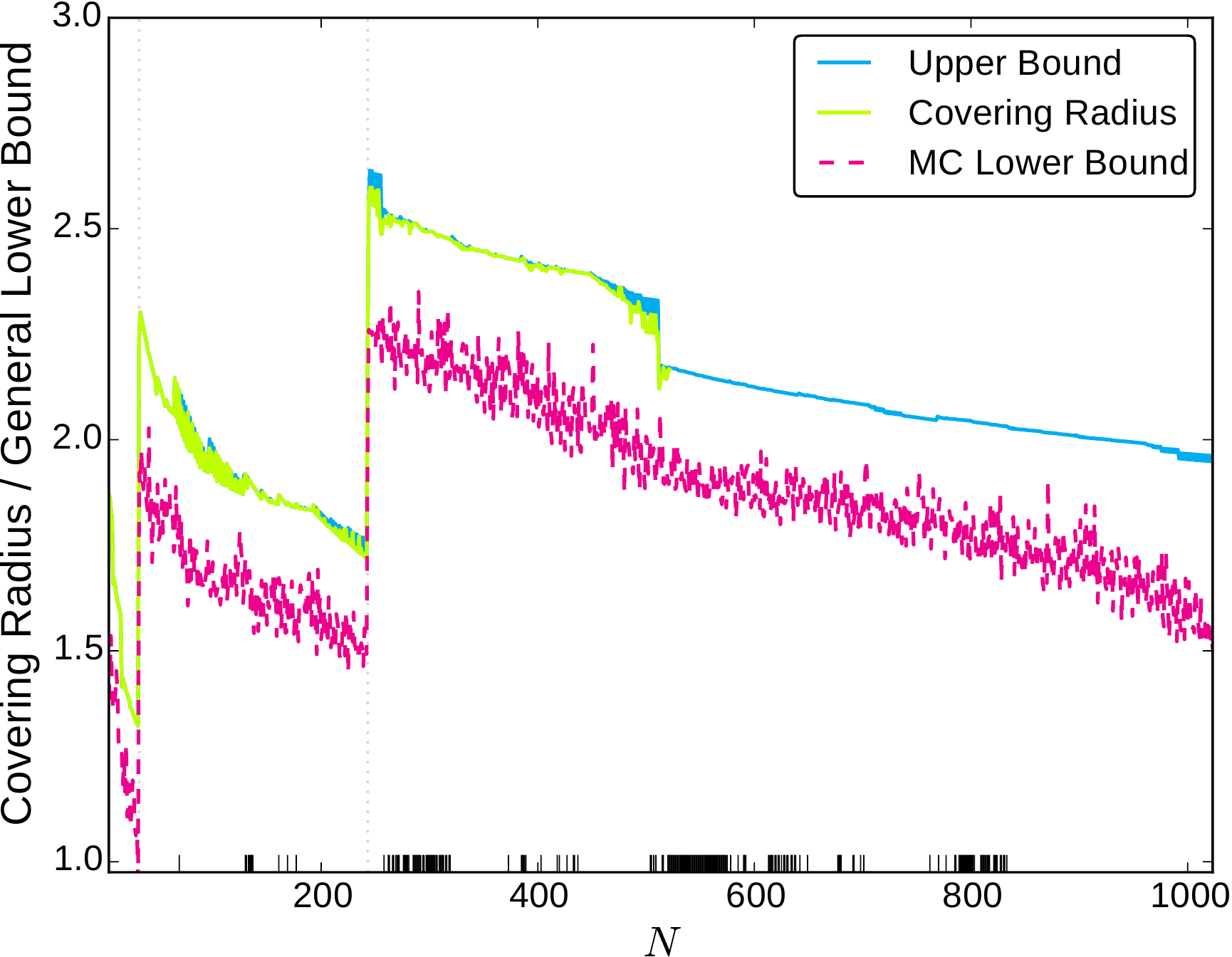}\label{fig:crbounds_vs_N10_dim5}}
\hfill\subfloat[$n = 10$]{\includegraphics[width=0.47\textwidth]{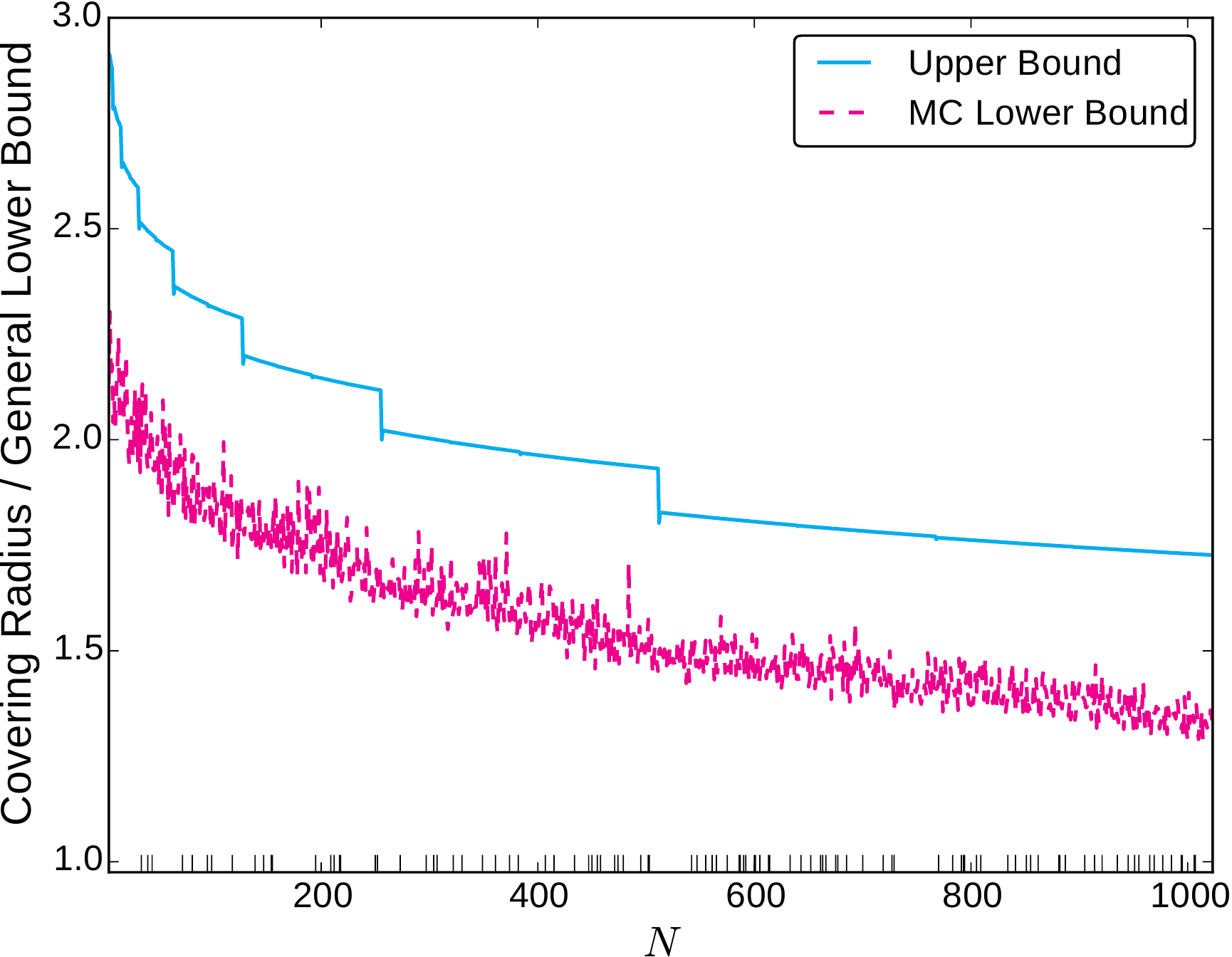}\label{fig:crbounds_vs_N10_dim10}}

%
%

\caption{Covering radius for the new stratified sampling algorithm with $b = \infty$. The event plots at the bottom of the subfigures indicate the cases where Alg.~\ref{alg:strat_sampling} without lines 6--7 obtains a lower covering radius than with lines 6--7.}
\label{fig:cr_bounds}
\end{figure*}

\section{Experimental Analysis}
\label{sec:experiments}

In this section, we analyze the produced point sets with raw summary characteristics, in a numerical integration setting, and as a method for worst-case optimization.
State of the art for numerical integration are quasirandom sequences \citep{Sobol1967,Faure2009}, and lattice rules \citep{Sloan1994}.
Thus, we compare stratified sampling to such methods, and more basic ones such as random uniform sampling and Latin hypercube sampling.

\subsection{Assessment with Summary Characteristics}

Before we come to the comparisons of sampling methods, we analyze the quality of upper and lower bounds for the covering radius in the context of stratified sampling. 
In Fig.~\ref{fig:cr_bounds}, we evaluate sets of up to 1024 points, generated by Alg.~\ref{alg:strat_sampling} (including lines 6--7). 
We plot the upper bound~\eqref{eq:cru}, a Monte Carlo lower bound using $M = 10N$ samples, and the exact value, as far as it can be computed in a timely manner. 
The exact algorithm as described in Sect.~\ref{sec:covering_radius} is realized with the Qhull library \citep{Barber1996}, via its SciPy interface.
There also exists a general lower bound for any covering radius of $N$ points in the $n$-dimensional unit hypercube, which is
\begin{equation}
\frac{1}{2\lfloor N^{1/n}\rfloor}\,,
\label{eq:general_lb}
\end{equation}
according to \citet{Sukharev1971}.
For a better visualization, our measured values are divided by this general lower bound~\eqref{eq:general_lb}, which is a step function that changes only at $n$-th powers of integers. 
These positions are indicated by vertical dotted gray lines.
The plots show that the upper bound is quite tight and often exact (29\%, 21\%, and 47\% of the cases where we computed the exact values in two, three, and five dimensions, respectively), while the lower bound with so few points is naturally quite noisy. 
However, already in this setting the lower bound is much more expensive than the upper bound, and its quality deteriorates with increasing dimension.
So we generally recommend the upper bound as the more useful and cheaper figure.

Event plots at the bottom of each subfigure indicate the cases where Alg.~\ref{alg:strat_sampling} without lines 6--7 achieves a lower covering radius. 
Interestingly, these cases are clustered together. 
Quite naturally, the avoidance of odd splits can only considerably reduce the covering radius for even numbers of points, which is the reason why we see high-frequency oscillations in the upper bound, especially in low dimensions.
With increasing dimension, however, this effect vanishes together with the usefulness of lines 6--7.
So, in ten dimensions the two algorithm variants almost always lead to the same covering radius. 
Table~\ref{tab:avoid_wins_losses} in the appendix contains the actual numbers of wins and losses for the two algorithm variants.

\begin{figure*}[p]
\centering

\subfloat[$n = 2$, $N = 100$]{\includegraphics[width=0.49\textwidth]{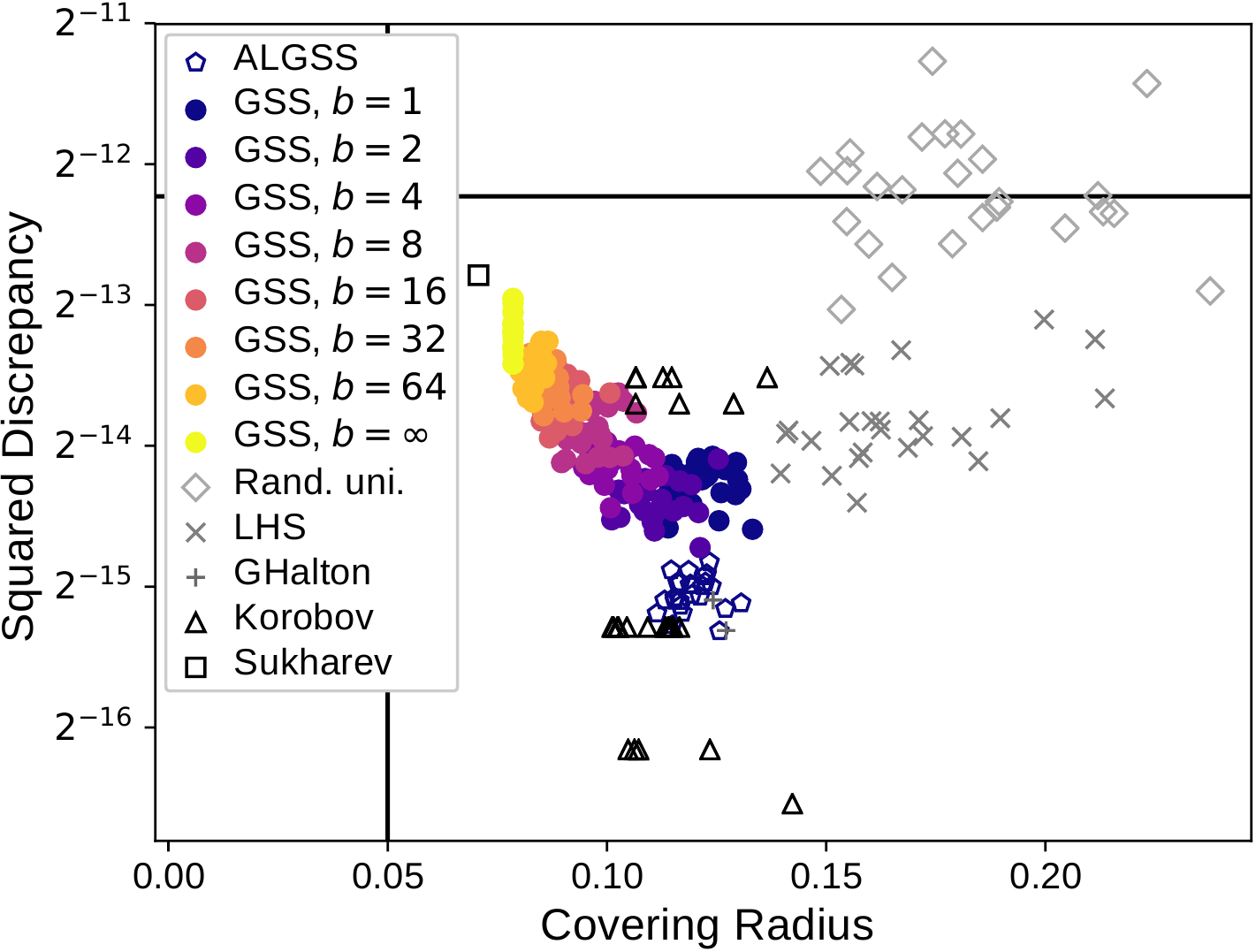}\label{fig:tradeoff2_100}}
\hfill\subfloat[$n = 2$, $N = 4900$]{\includegraphics[width=0.49\textwidth]{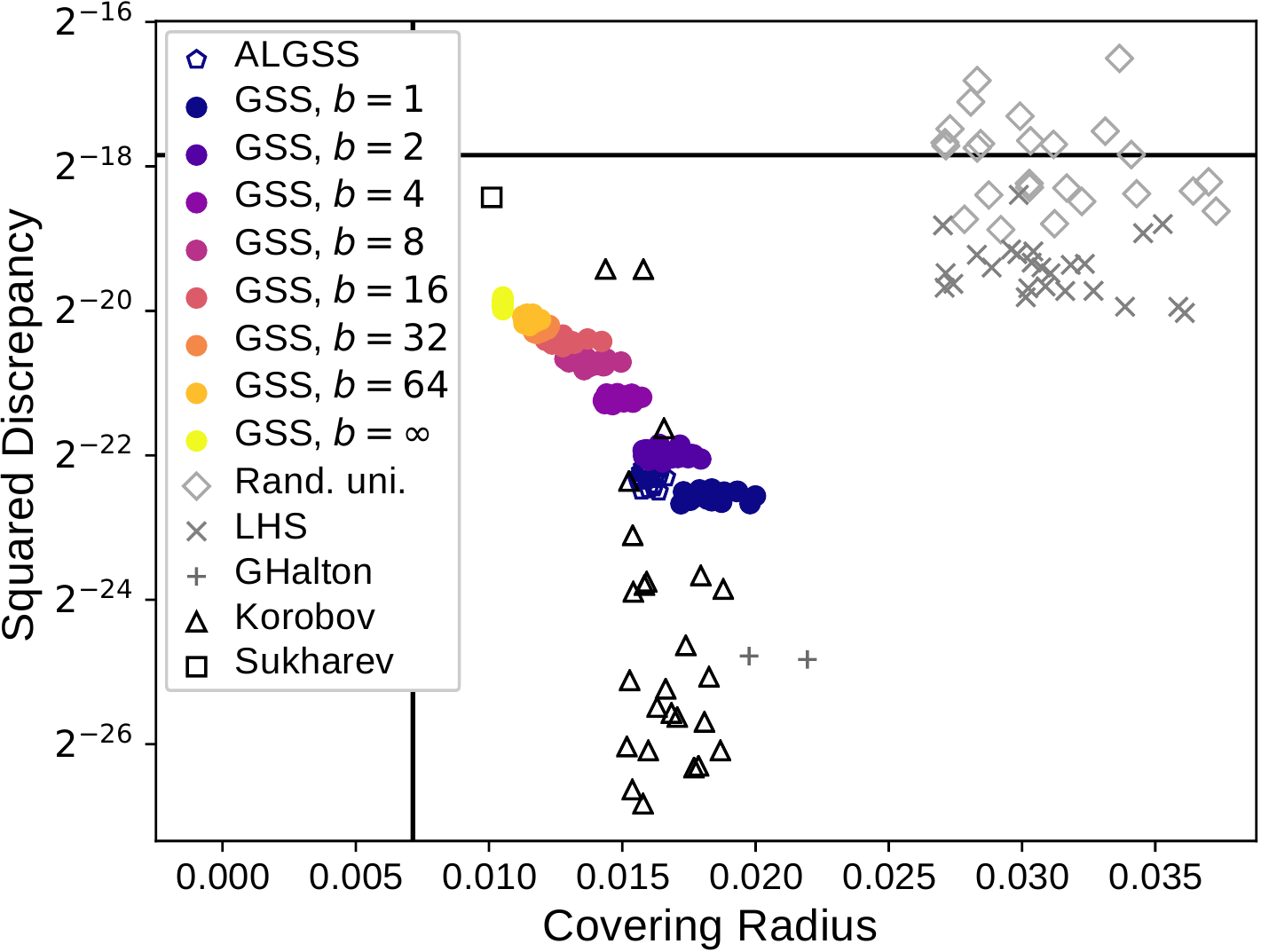}\label{fig:tradeoff2_4900}}

\subfloat[$n = 5$, $N = 100$]{\includegraphics[width=0.49\textwidth]{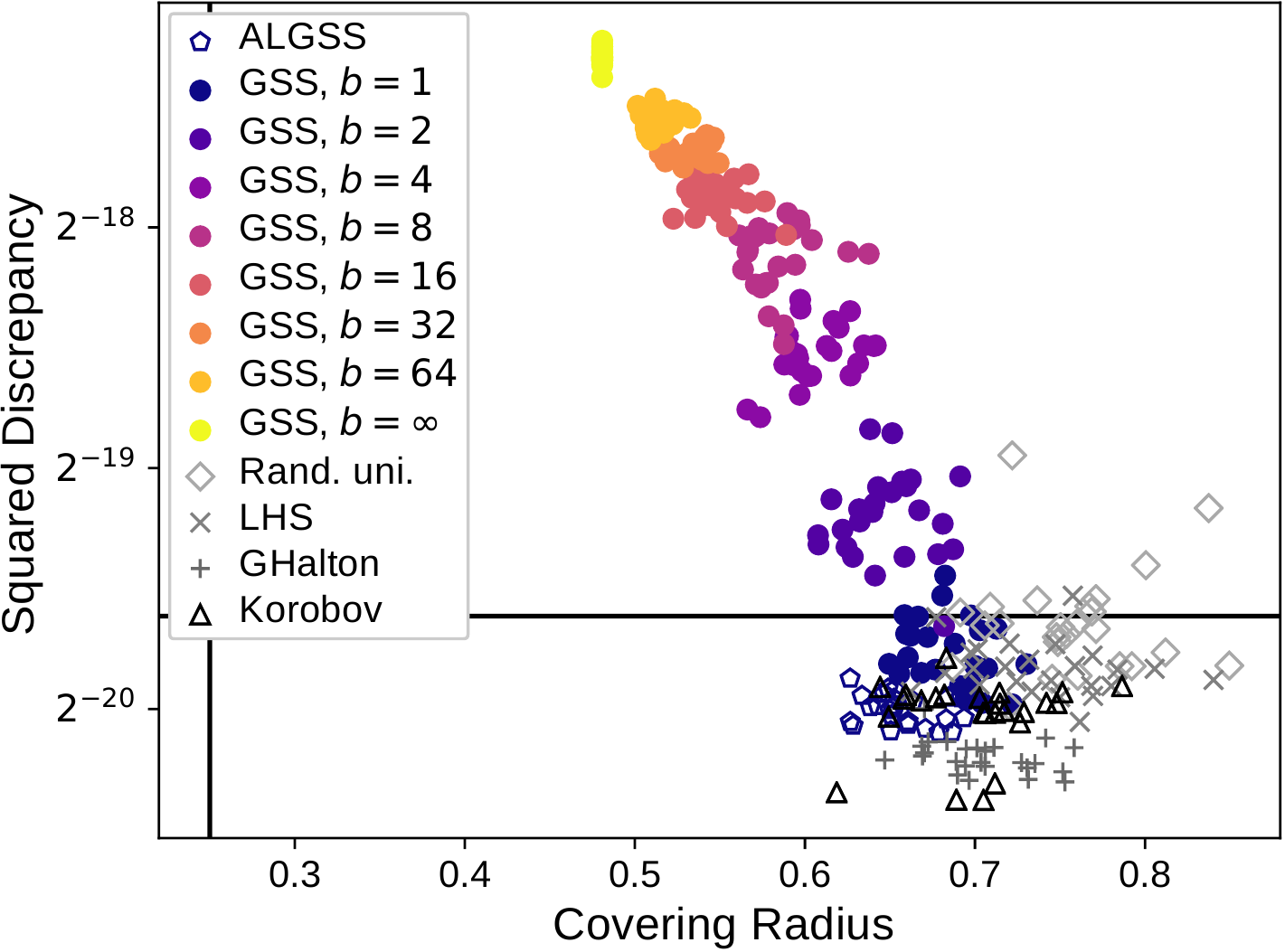}\label{fig:tradeoff5_100}}
\hfill\subfloat[$n = 5$, $N = 4900$]{\includegraphics[width=0.49\textwidth]{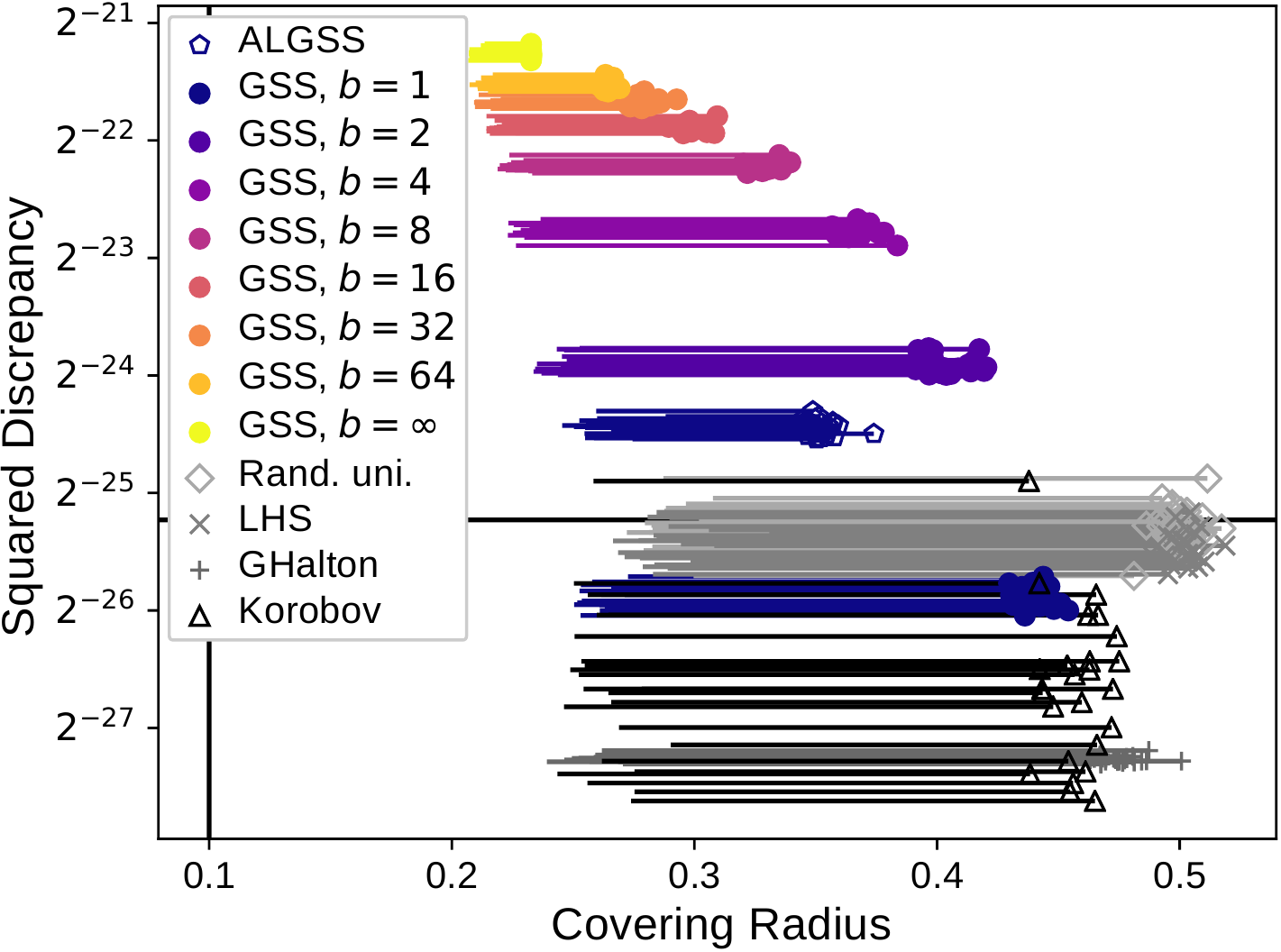}\label{fig:tradeoff5_4900}}

\subfloat[$n = 10$, $N = 100$]{\includegraphics[width=0.496\textwidth]{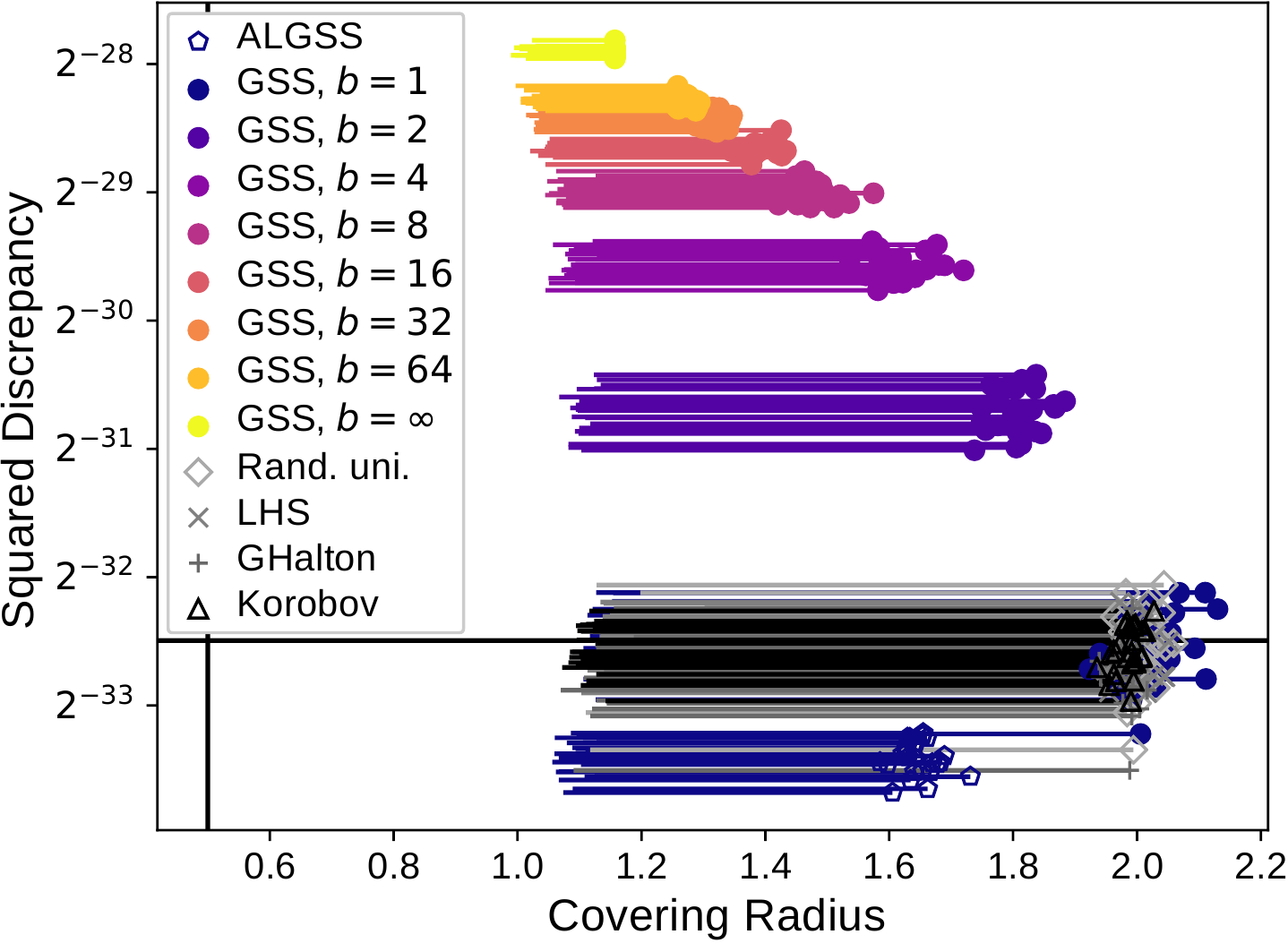}\label{fig:tradeoff10_100}}
\hfill\subfloat[$n = 10$, $N = 4900$]{\includegraphics[width=0.49\textwidth]{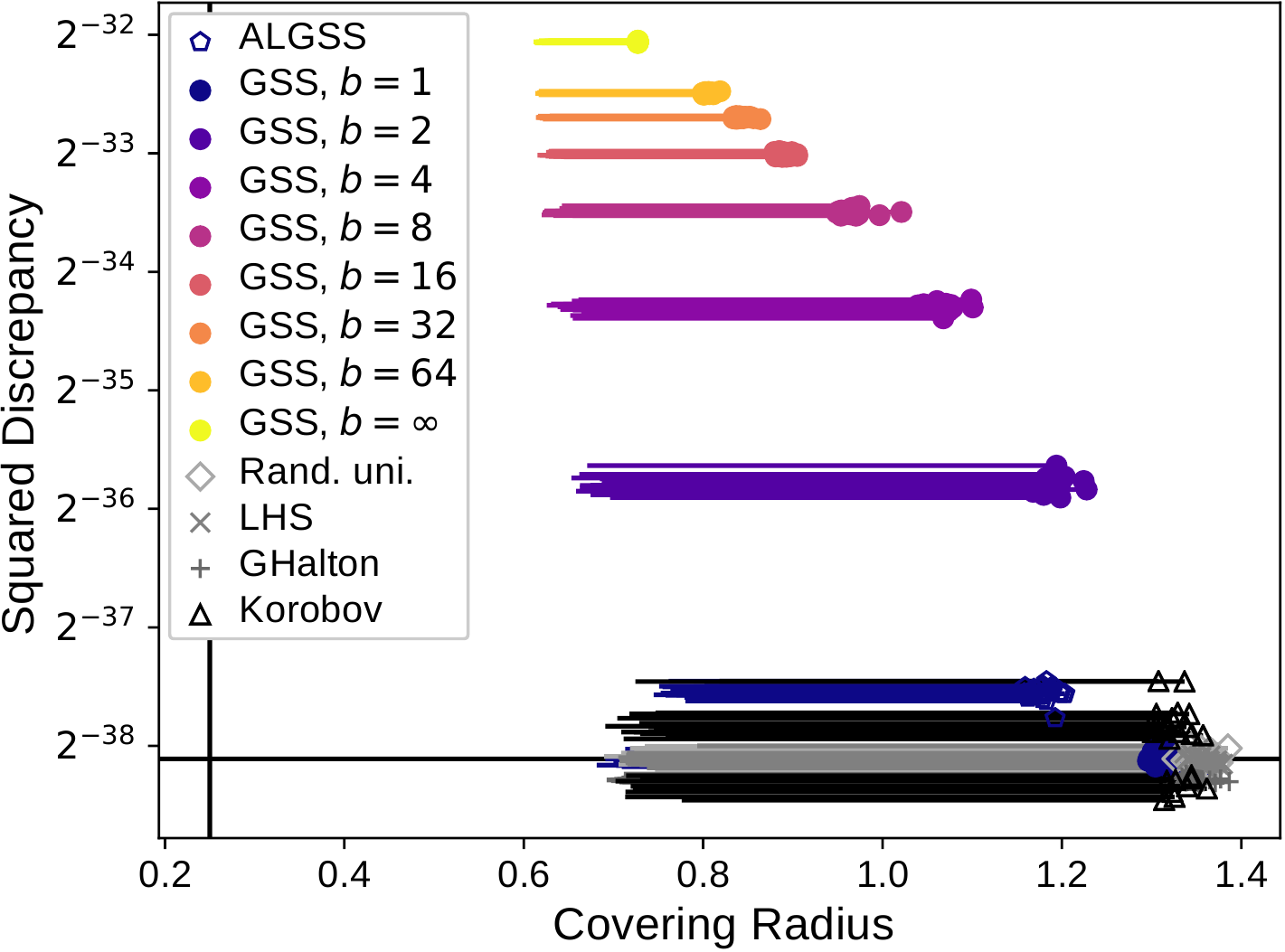}\label{fig:tradeoff10_4900}}

%
%
%
%

\caption{Discrepancy and covering radius of different sampling algorithms. 
In case of approximate covering radius values, the marker is drawn at the upper bound and an error bar extends to the lower bound. For both objectives lower is better.
}
\label{fig:tradeoff_cr_disc}
\end{figure*}

\begin{table*}[t]
\caption{Overview of sampling algorithms appearing in the experiments.}
\centering
\begin{tabularx}{\textwidth}{lLl}
\toprule
Abbreviation & Algorithm & Reference\\
\midrule
SRS & Simple random sampling & \\
LHS & Latin hypercube sampling & \cite{McKay1979}\\
GHalton & Generalized Halton sequence & \cite{Faure2009}\\
Korobov & Randomly shifted Korobov lattice & \cite{Sloan1994}\\
LKorobov & Randomly shifted Latinized Korobov lattice & \cite{Sloan1994}\\
PSS-$x^y$ & Partially stratified sampling, created by padding $y$ $x$-dimensional stratified samples & \cite{Shields2016}\\
LPSS & Latinized PSS & \cite{Shields2016}\\
GSS & Generalized stratified sampling & here\\
ALGSS & Approximately latinized GSS & here\\
LGSS & Exactly latinized GSS & here\\
ALGPSS & PSS, where the lower-dimensional samples are created with ALGSS & here\\
LGPSS & PSS, where the lower-dimensional samples are created with LGSS & here\\
\bottomrule
\end{tabularx}
\label{tab:overview_algos}
\end{table*}

Asymptotically, every low-discrepancy point set also has a low covering radius (but not vice versa)~\cite[p.~152]{Niederreiter1992}. 
However, for concrete point sets, quite a trade-off can be detected between the two, as our experiments show.
This is illustrated in Fig.~\ref{fig:tradeoff_cr_disc}, where we investigate sets of sizes 100 and 4900 in dimensions 2, 5, and~10. 
Results of stratified sampling with different values of $b$ are shown as colored dots.
LGSS is not included in these figures because of its higher runtime.
For a comparison, we add random uniform points (SRS), Latin hypercube designs, generalized Halton sequences\footnote{Implemented in the ghalton Python library at \url{https://github.com/fmder/ghalton} \citep{DeRainville2012}. 
}, a Korobov lattice, and, where possible, a Sukharev grid.
An overview of the acronyms used in this experiment and the following ones is given in Tab.~\ref{tab:overview_algos}.
Each configuration is sampled 25 times. 
(For the generalized Halton sequence, only two possible permutations of the involved prime numbers exist in two dimensions.)
The Korobov lattices in Fig.~\ref{fig:tradeoff_cr_disc} are the best of~30 randomly chosen lattices according to their separation distance. 
This maximization of the minimal distance between points in the lattice is proposed by 
\cite{Dammertz2008}, and the number 30 is the default setting in the Lattice Builder software~\citep{Lecuyer2016}.
Additionally, the Korobov lattices are randomly shifted in the sense of~\cite{Cranley1976}.
It is also very easy to generate Korobov lattices with LH property, but in this figure, we do not yet enforce this restriction, because we want to observe the whole distribution.
However, some of them are Latin by chance.

The vertical lines in Fig.~\ref{fig:tradeoff_cr_disc} mark~\eqref{eq:general_lb}, the horizontal lines mark~\eqref{eq:expected_value_disc}.
Where it is feasible we plot the exact covering radius, otherwise we use the upper bound~\eqref{eq:cru} and~a Monte Carlo lower bound with $M = 10^4 \cdot 2n$. 
For the point sets that do not result from stratified sampling, we first have to find a partition of the unit hypercube to compute the upper bound for the covering radius. 
We do this with a variation of the mean-split algorithm~\citep{Wu1985}, which needs $O(nN \log_2 N)$ average runtime.
This algorithm variant chooses the longest side of a hyperbox for a split, as the original (ties are broken randomly).
However, the heuristic for the determination of the split position differs. 
Ours works as follows: we calculate the mean value of the points in the chosen dimension as a preliminary split position. 
Then, we identify the closest neighbor points to this preliminary position.
The actual position is then chosen as the mean of these two points. 
Once we have partitioned the whole space, with each hyperbox containing one point, the upper bound can be computed just as for stratified sampling.
We iterate this process 10 times, due to the stochasticity in the partitioning, and take the best upper bound we find.

\begin{figure*}[t!]
\centering
\subfloat[ $n = 2$]{\includegraphics[scale=0.5]{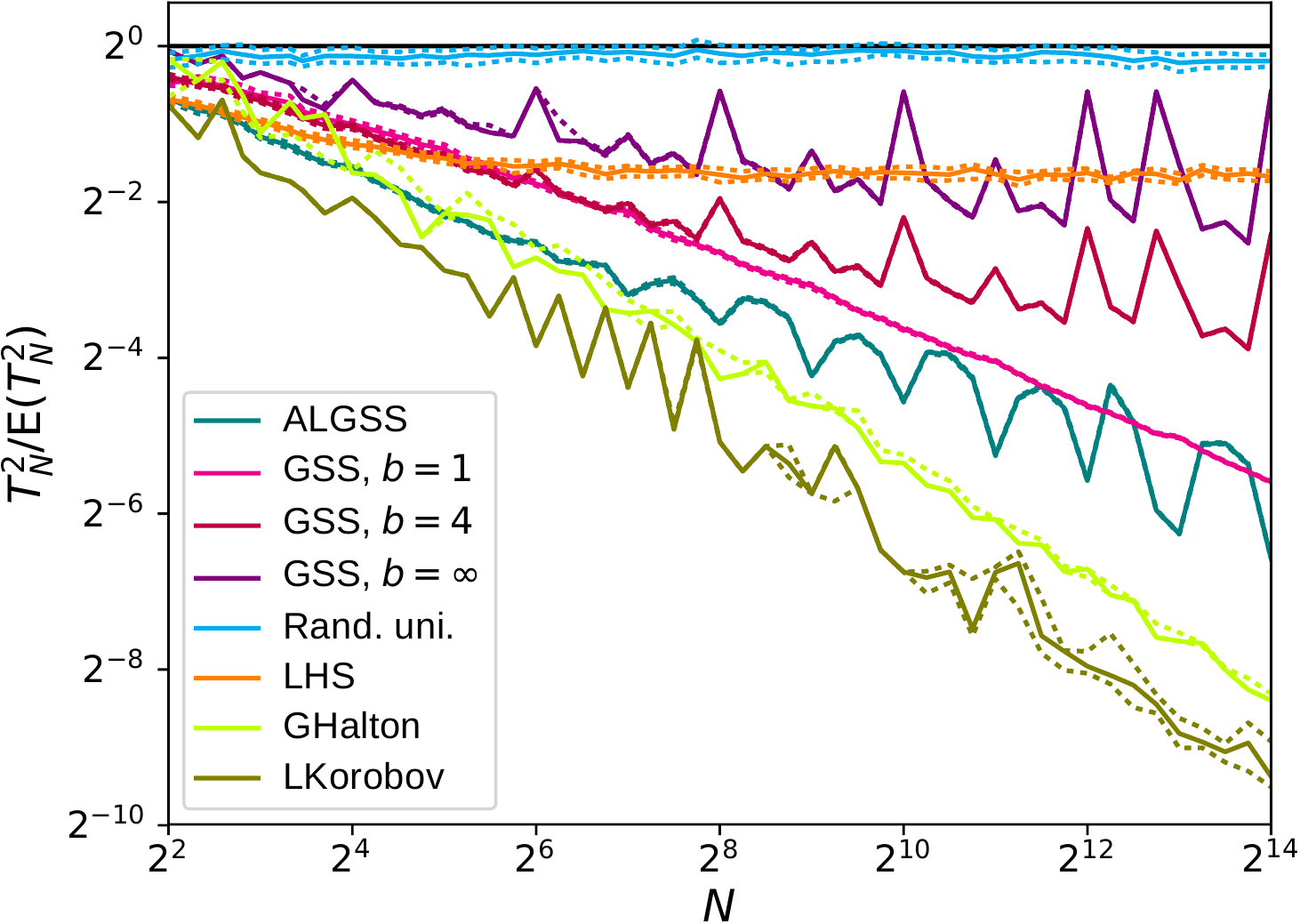}%
\hfill\includegraphics[scale=0.5]{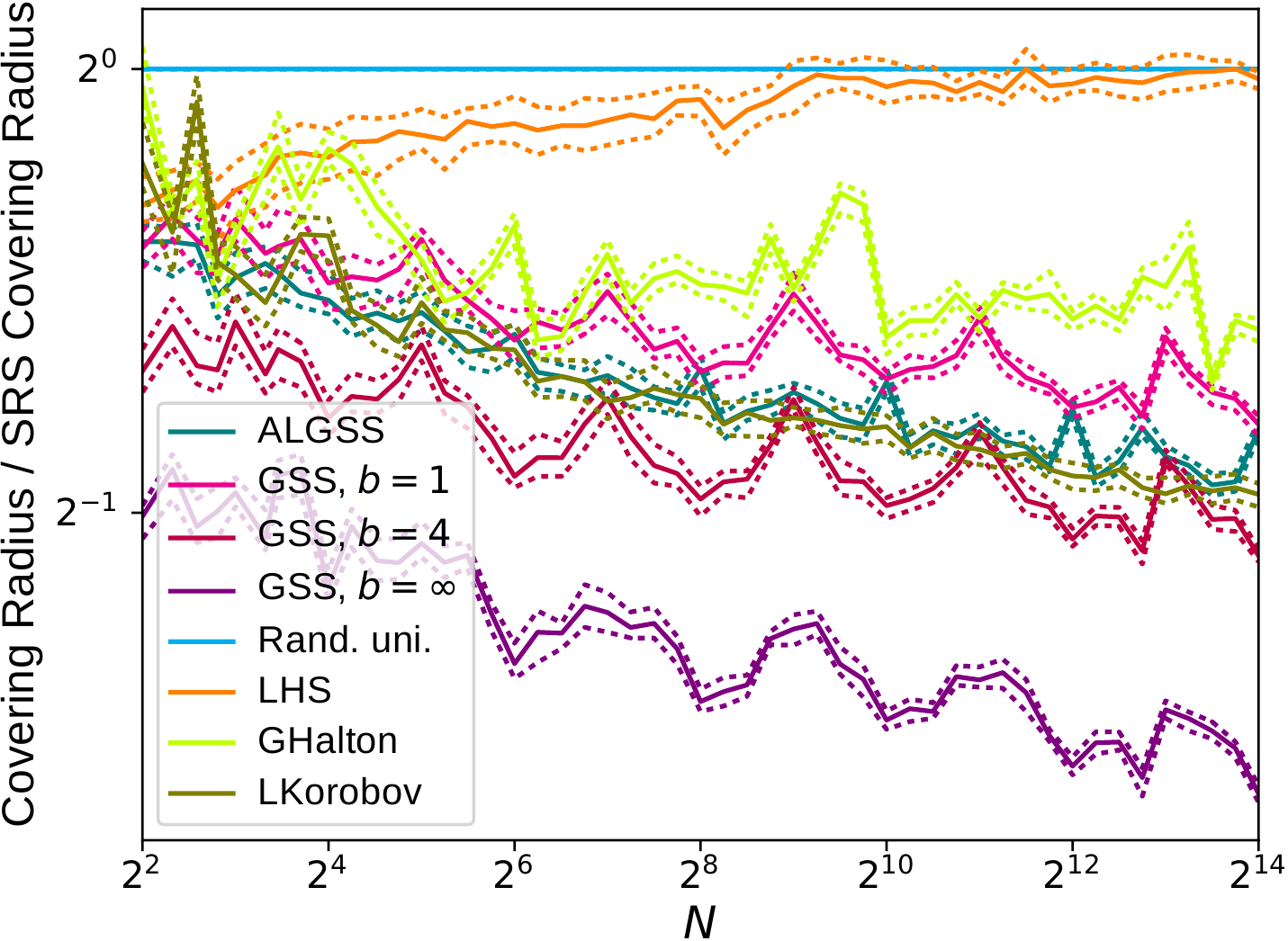}\label{fig:convergence2}}

\subfloat[ $n = 3$]{\includegraphics[scale=0.5]{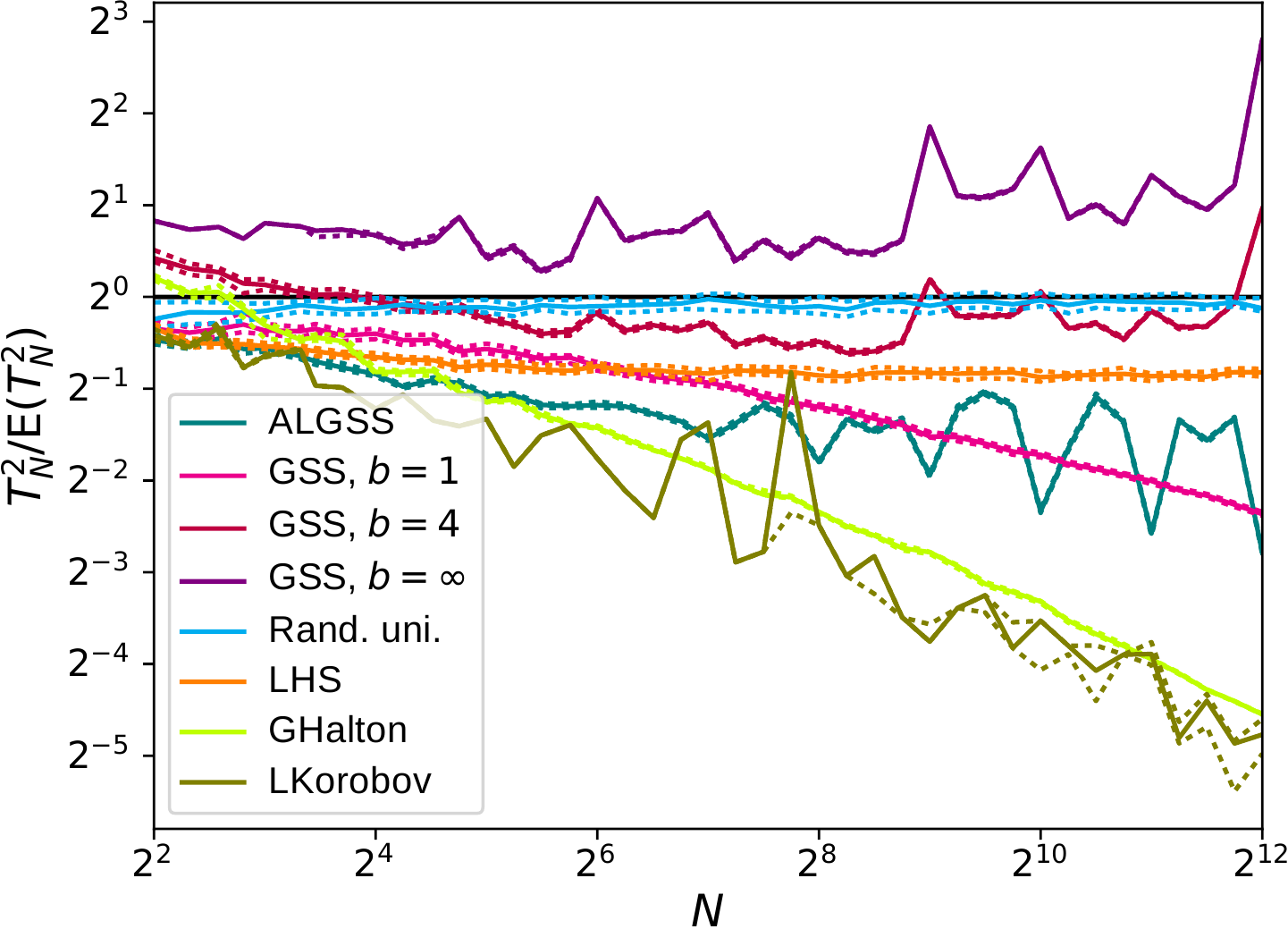}%
\hfill\includegraphics[scale=0.5]{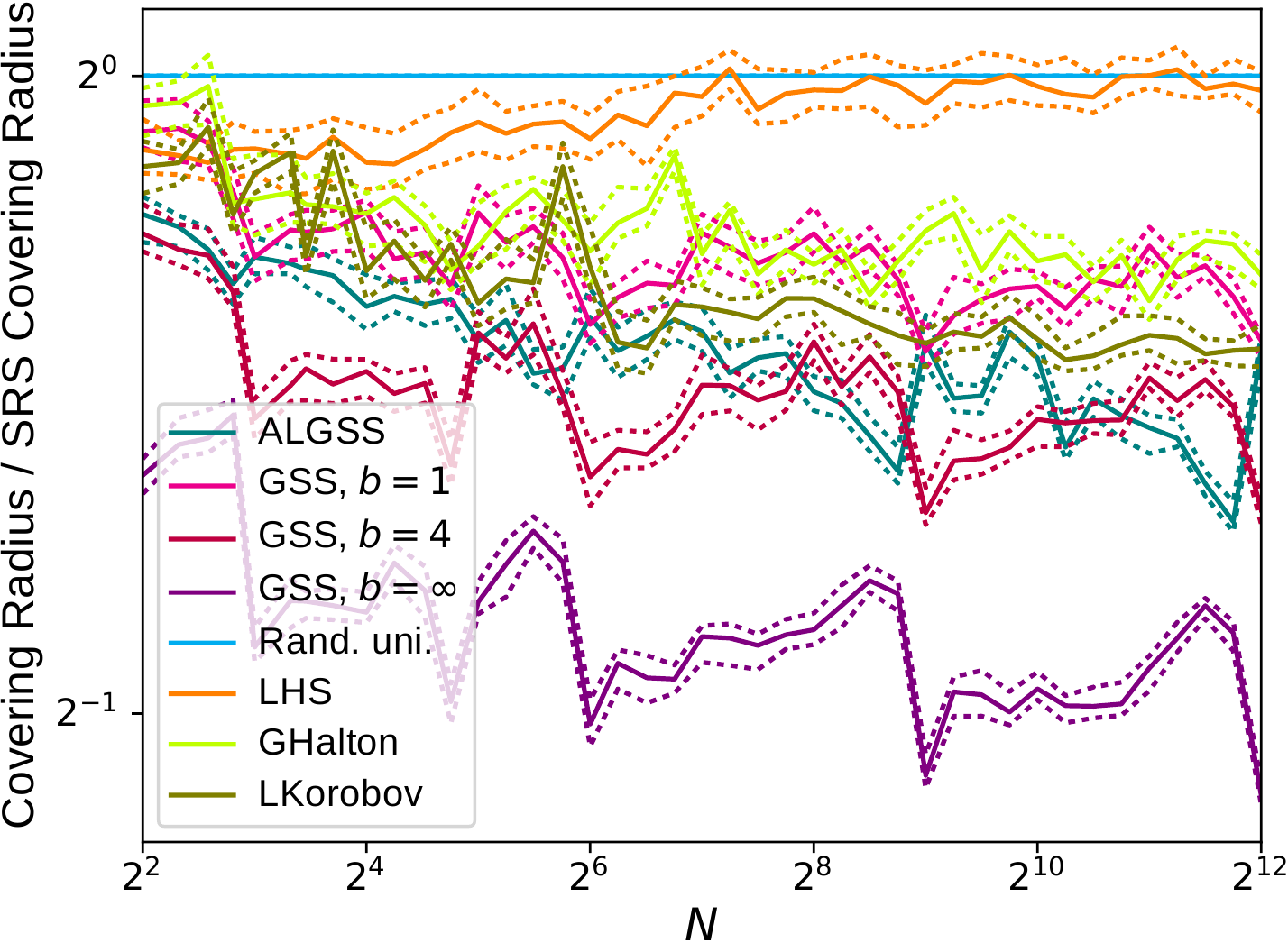}\label{fig:convergence3}}

%
%

\caption{Discrepancy and covering radius of various point sets, depending on sample size. 
The horizontal black line marks the expected discrepancy for random uniform points.
}
\label{fig:convergence_cr_disc}
\end{figure*}

Figure~\ref{fig:tradeoff_cr_disc} is to be read in a Pareto sense, i.\,e., we are interested in points for which there is no other point which is better in every objective.
Except for SRS, LHS, GSS with $b = 1$, and GHalton, all methods are clearly non-dominated in some of the scenarios.
Especially in low dimensions, i.\,e., at least for $n = 2$, it is easily possible to keep both discrepancy and covering radius below the values of random uniform points.
In higher dimensions, larger sample sizes would be required to obtain the same effect. 
Otherwise, the difference between stratified sampling with $b = 1$ and random uniform sampling vanishes for the investigated sample sizes.
However, it is always possible to improve the covering radius, albeit at the cost of a worse discrepancy.
The Sukharev grid and the Korobov lattice represent the two extremes among the contestants.
The plots show that stratified sampling variants fills a gap between them.
ALGSS is an interesting case because it seems to excel especially in cases with high dimensions and low numbers of points.

Figure~\ref{fig:convergence_cr_disc} gives another view of the trade-off between discrepancy and covering radius in two and three dimensions. 
Here, we plot the measures against the sample size, with $\lfloor400/n\rfloor$ stochastic replications to filter out the noise. 
Solid lines mark the median and dotted lines pointwise 95\% confidence intervals for the median. 
The figure confirms our previous observations:
we can see that the Korobov lattice (now and for the remainder of the paper restricted to Latin instances) has the best convergence order for discrepancy.
As already seen in Fig.~\ref{fig:tradeoff_cr_disc}, the discrepancy of stratified sampling with $b \gg 1$ is worse than with random uniform sampling (for $n \geq 3$ and $N \leq 2^{12}$).
GSS with $b = \infty$ is the best in terms of covering radius, but the worst in discrepancy.
The quasirandom methods, on the other hand, still have a medium covering radius, despite their good discrepancy, in accordance with theory~\cite[p.~152]{Niederreiter1992}.
LHS acts very similarly to random uniform sampling, and stratified sampling with $b > 1$ exhibits spikes at $N = 2^{ni}$, $i \in \mathbb{N}$, corresponding to the cases where the conventional stratified sampling is recovered.

\subsection{Numerical Integration}

To verify the utility of our latinized GSS variants, we replicate and extend some experiments from~\cite{Shields2016}. 
The task in these experiments is to estimate integrals on two artificial test functions and one plate buckling problem.
The number of points is $N = 625$ in all the cases here and we use the standard deviation of the obtained mean estimates over 5000 replications as the performance measure.
The two artificial functions are Rosenbrock's function and Schwefel's double-sum function, both in 100 dimensions.
Rosenbrock's function is sampled in the unit hypercube, while for the double-sum, the points are transformed to two normal distributions with means zero and one, respectively.
To have a challenging competitor, the exact optimum regarding separation distance is determined for the Korobov lattice here.

Table~\ref{tab:artificial_integration} shows the results of this experiment.
On the Rosenbrock function a partially stratified sampling design remains the best approach.
The two latinized GSS variants only obtain results similar to the LHS, probably due to the high dimensionality.
Therefore, we also test combinations of LPSS and (A)LGSS, which turn out to be competitive to the conventionally stratified PSS variants.
LKorobov and ALGSS are the best methods for one double-sum scenario, respectively.

Table~\ref{tab:plate_buckling} contains the results of the plate buckling experiment, which is concerned with the structural strength of steel plates.
For a detailed description, we refer to the original paper.
This problem is only six-dimensional, and here LGSS obtains the best performance, followed by ALGSS.
Note that our mean estimates slightly differ from the ones published by~\cite{Shields2016}.
This is explained by a small code change in their supplementary material since the publication of the paper\footnote{This is noted in the file ``PSS\_Plate\_Buckling.m'' at \url{https://www.mathworks.com/matlabcentral/fileexchange/54841-partially-stratified-sampling}.}.
The mean estimate of the GHalton sample is noticably off here.

We conclude that the generalized stratification does not impair the performance in these estimation scenarios.
It is easier to work with than partial stratification, because it does not require problem knowledge, as PSS does.
But it can benefit from it just as well.
Furthermore, note that similarly to Fig.~\ref{fig:tradeoff_cr_disc}, there is no method that wins every considered scenario.

\begin{table}[t!]
\caption{Standard deviation of mean value estimates for the Rosenbrock and double-sum functions. Best results are in bold.}
\centering
\sisetup{
table-number-alignment = center,
table-figures-integer  = 3,
table-figures-decimal  = 3
}
\begin{tabularx}{\columnwidth}{
L
S[table-format=1.3,table-auto-round]
S[table-format=3.1,table-auto-round,table-figures-decimal=1]
S[table-format=4.1,table-auto-round]
}
\toprule
Design & {Rosenbrock} & {Double-sum} & {Double-sum}\\
 & & {$N(0, 1)$} & {$N(1, 1)$} \\
\midrule
SRS & 8.79469895572 & 232.971950961 & 2955.71294186\\
LHS & 6.69622775781 & 232.918119283 & 238.174098015\\
GHalton & 6.52896648066 & 188.870550645 & 781.419202619\\
LKorobov & 4.21733190503 & {\textbf{185.5}} & 316.920223033\\
PSS-$2^{50}$ & 4.82117728035 & 225.288747987 & 401.569809622\\
PSS-$4^{25}$ & 4.65589541084 & 227.341719426 & 963.67092684\\
LPSS-$2^{50}$ & 4.86086189811 & 225.63387933 & 231.194061833\\
LPSS-$4^{25}$ & {\textbf{3.842}} & 226.727987031 & 231.71918833\\
GSS & 8.76664186341 & 232.945745151 & 2820.82988337\\
ALGSS & 6.7847431978 & 222.469846388 & {\phantom{0}\textbf{228.0}}\\
LGSS & 6.87371578037 & 232.471228439 & 237.82505727\\
ALGPSS-$2^{50}$ & 4.87754448337 & 227.613371911 & 233.370698223\\
ALGPSS-$4^{25}$ & 4.94405791158 & 230.395233388 & 235.588618556\\
LGPSS-$2^{50}$ & 4.80234079139 & 230.330667791 & 236.490975006\\
LGPSS-$4^{25}$ & 3.98393633143 & 227.871072169 & 232.205013565\\
\bottomrule
\end{tabularx}
\label{tab:artificial_integration}
\end{table}

\begin{table}[t!]
\caption{Results of the plate buckling strength experiment.}
\centering
\sisetup{
table-number-alignment = center,
table-figures-integer  = 1,
table-figures-decimal  = 3
}
\begin{tabularx}{\columnwidth}{
L
S[table-format=1.5,table-auto-round,table-figures-decimal=5]
S[table-format=1.2e-1,table-auto-round,scientific-notation=true]
}
\toprule
Design & {Mean strength $E(\phi)$} & {Std. Dev. of $E(\phi)$}\\
\midrule
SRS & 0.585762484749 & 0.00104065837465 \\
LHS & 0.585758679954 & 8.00481483176e-05 \\
GHalton & 0.585882185895 & 0.000119424164369 \\
LKorobov & 0.585758840199 & 0.000127931604333 \\
PSS-$2^3$ & 0.585753895775 & 0.000133963456205 \\
PSS-$2^2 1^2$ & 0.585755549112 & 0.000120721842853 \\
PSS-$4^1 1^2$ & 0.585751126548 & 0.00029421101053 \\
LPSS-$2^3$ & 0.585758385058 & 7.15636617678e-05 \\
LPSS-$2^2 1^2$ & 0.585757772984 & 7.36239235769e-05 \\
LPSS-$4^1 1^2$ & 0.585759443007 & 6.13868140918e-05 \\
GSS & 0.585755549167 & 0.000521031277082 \\
ALGSS & 0.585760069444 & 6.10233712891e-05 \\
LGSS & 0.585757829911 & {$\mathbf{5.71 \times 10^{-5}}$} \\
\bottomrule
\end{tabularx}
\label{tab:plate_buckling}
\end{table}

\subsection{Worst-case Optimization}

In optimization, we may use $\hat{f}^* = f(\hat{\vec{x}}^*)$, $\hat{\vec{x}}^* = \operatorname{\arg\min}\{f(\vec{x}) \mid \vec{x} \in P\}$, as an estimate of the global minimum $f(\vec{x}^*)$ of a function~$f: X \to \mathbb{R}$.
Again, $P \subset X$ is a finite approximation set. 
For the estimator $\hat{f}^*$ it holds that $\hat{f}^* - f(\vec{x}^*) \leq \omega(f, d_\mathrm{cr}(P, X))$, where 
$$\omega(f, t) = \underset{\|\vec{x}_i - \vec{x}_j\|_2 \leq t}{\sup_{\vec{x}_i,\vec{x}_j \in X}} \{|f(\vec{x}_i) - f(\vec{x}_j)| \}$$
is, for $t \geq 0$, the modulus of continuity of $f$~\cite[p.~149]{Niederreiter1992}.
So, the worst-case error of our estimator is bounded by a function depending on the covering radius of~$P$ and thus, it seems advisable to choose point sets with low covering radius. 
In this experiment we want to analyze if this theoretical guideline can be rediscovered in the measurements.

The extreme case where no local search is performed and the whole budget is spent on a uniformly distributed set~$P$ represents the most conservative approach possible. 
Such an approach is, e.\,g., taken by~\cite{Yakowitz2000}.
Uniform sampling can also appear as a component of more sophisticated algorithms for global optimization, e.\,g., in the simplest case a single local search could be started from $\hat{\vec{x}}^*$ \cite[p.~66]{Toern1989}.
For our purposes it suffices to assess the quality of the uniform point sets with the aforementioned measure~$\hat{f}^* - f(\vec{x}^*)$, because a better quality here would also give any following local search a head start.

\begin{figure*}[p]
\centering
\includegraphics[width=0.975\textwidth,clip,trim=0 20 0 0]{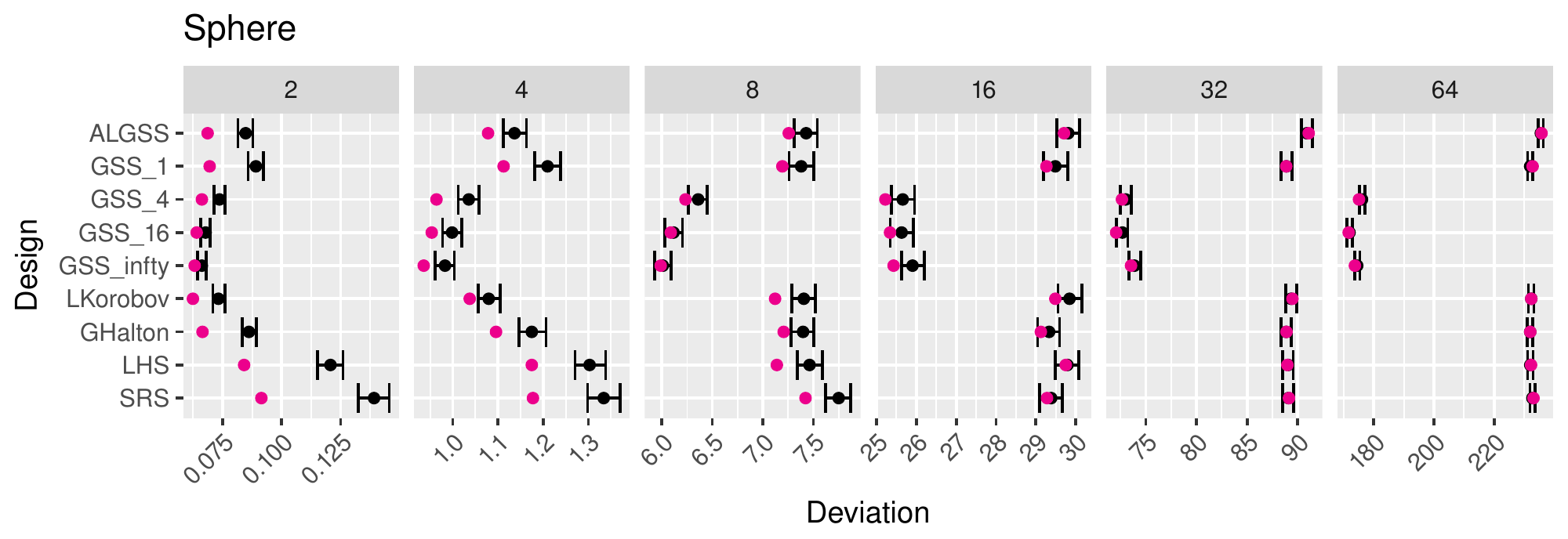}
\includegraphics[width=0.975\textwidth,clip,trim=0 20 0 0]{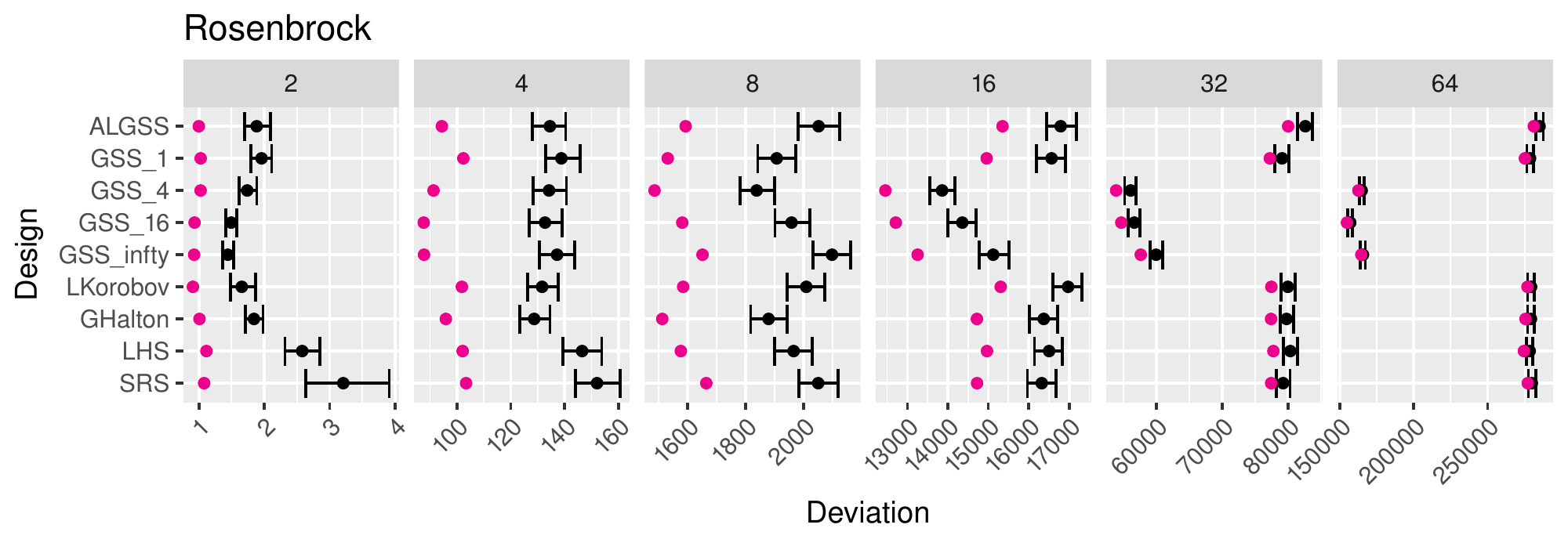}
\includegraphics[width=0.975\textwidth,clip,trim=0 20 0 0]{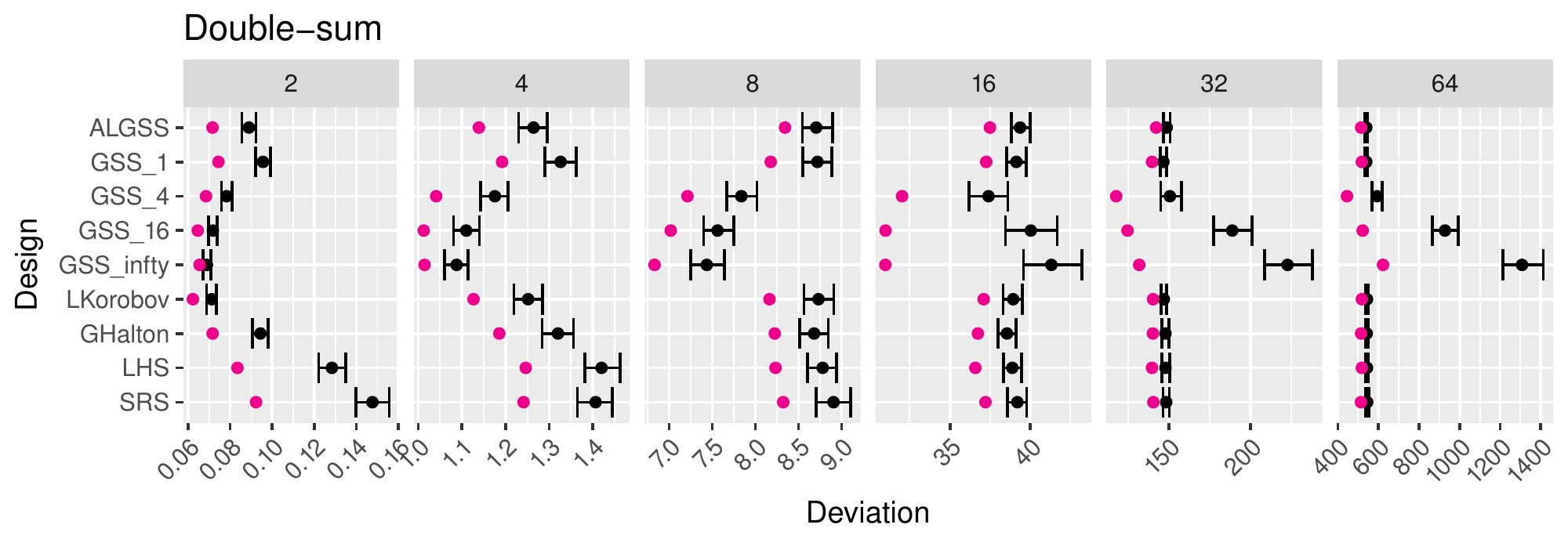}
\includegraphics[width=0.975\textwidth]{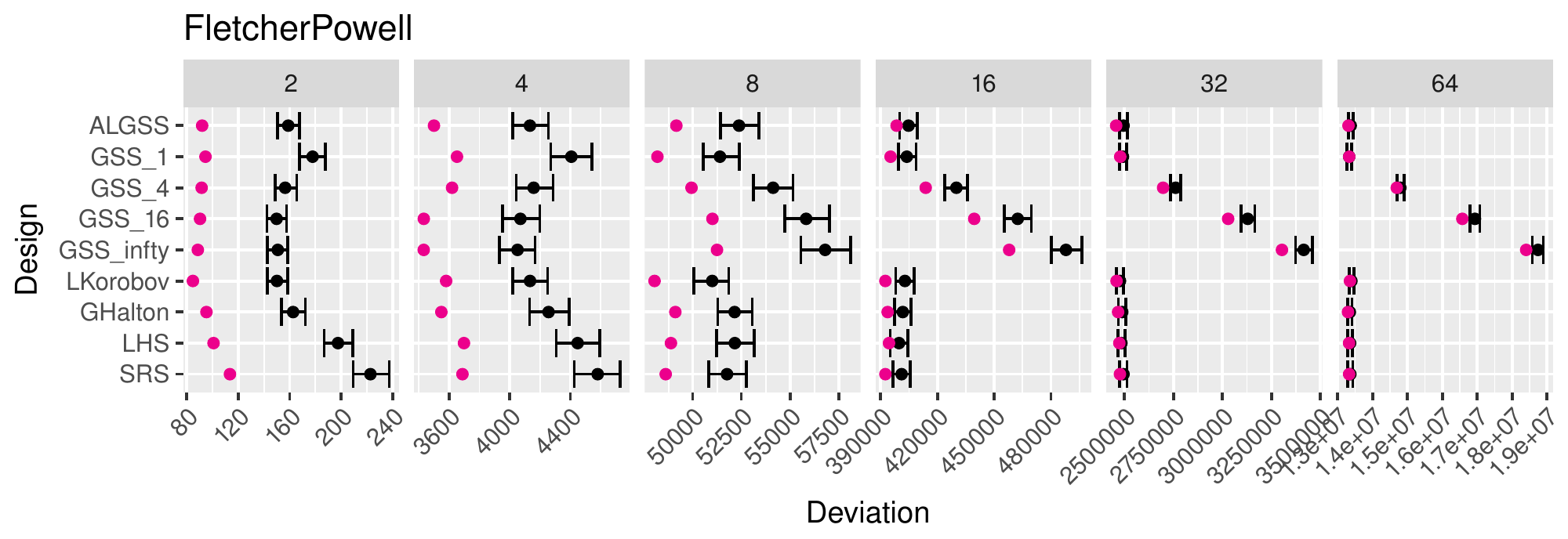}

%
%
%
\caption{Approximation errors to the global optimum, depending on sampling methods and dimension. 
Error bars denote mean values with bootstrapped 95\% confidence intervals, magenta dots mark the median.
}
\label{fig:global_opt}
\end{figure*}

We are employing sets of size $50n$ in dimensions $n = 2, 4, 8, 16, 32, 64$ for four different optimization test functions.
$50n$ is a common number of points, i.\,e., it is used in recent clustering methods for global optimization~\cite[p.~154]{Preuss2015}.
Next to the previously used Rosenbrock and double-sum functions, we additionally try the sphere function $f(\vec{x}) = \vec{x}^\top\vec{x}$ and the generator of~\cite{FlePow63} (FP in the following), which produces random instances of multimodal problems with $2^n$ optima.
The latter one has a search space of $[-\pi, \pi]^n$, so out of convenience we scale our point sets to this region for all functions.
The other three functions are randomly shifted so that their global optimum is distributed uniformly in $[-\pi, \pi]^n$. 
Note that the volume of the search space grows exponentially, while the number of points only increases linearly, and the function values are not normalized.
Thus, the measured deviation must necessarily increase with dimension.
Figure~\ref{fig:global_opt} shows the results of 2000 replications of this experiment.
From the performance of GSS with $b = \infty$, we can deduce that minimizing the covering radius as much as possible can be beneficial at least in low dimensions. 
For $n \leq 4$, GSS even regularly beats the Korobov lattice.
The Korobov lattices are constructed as the best of 1000 randomly generated ones here, while it is ensured that for $N \leq 1000$ (and thus for $n \leq 16$) the optimal lattice is found.
So the Korobov lattices used here should be the best possible choice in terms of discrepancy.
In high dimensions the performance of GSS is either extremely good (on sphere and Rosenbrock's problem), or extremely bad (double-sum and FP). 
This indicates that while in some circumstances a low covering radius is more important for a good performance, in other circumstances a low discrepancy is mandatory. 
From the differences between mean and median, it can also be observed that the distributions for GSS with $b > 1$ are more skewed than the other ones.
A partial explanation for these observations may be ill-conditioning of the functions, which would probably require good distributions only in certain lower-dimensional subspaces.
\cite{Keller2006} mentions a similar problem in computer graphics, when the integration domain is much different from the unit hypercube.

\section{Conclusion}
\label{sec:conclusion}

We presented a generalized stratified sampling algorithm, which uses binary space partitioning to enable an arbitrary number of points to be sampled. 
Uniformity is guaranteed by ensuring that the final partition of the space consists of equal-sized hyperboxes.
The algorithm is easier to work with than other stratified sampling approaches and can also be combined with latinization to improve the discrepancy of the produced point sets.
Furthermore, we proposed to alternatively sample each hyperbox with the Bates distribution.
By varying the distribution's parameter, a changeover between low discrepancy and low covering radius can be obtained. 
Regarding these two measures, the resulting samples fill a gap between the Sukharev grid and quasirandom sequences. 
Additionally, we demonstrated how to compute lower and upper bounds for the covering radius, because the exact computation has a prohibitive time complexity in high dimensions due to a necessary Voronoi tessellation. 
The upper bound is important for minimizing the covering radius. 
It only needs linear time, reuses the hyperboxes of the stratified sampling or can also be obtained for arbitrary point sets, if we build a partition of the space around the points retroactively.

Our experiments with problems from numerical integration and worst-case optimization affirm the competitiveness of the proposed generalized stratified sampling variants and show several occasions where they beat state-of-the-art methods. 
As is inevitable in practice, the best method varies slight\-ly, depending on the dimension, the number of points sampled, and other problem properties. 
But so far we can say that both the Korobov lattice and (A)LGSS are good general purpose methods.
The speed of both algorithms can be traded off against quality to some extent, so a user's preference may actually depend on the implemented variant.
Finally, the minimization of the covering radius for optimization seems an interesting topic.
In the future, further research might work out more clearly when it is safe to rely on covering radius only for optimization performance.

\begin{table*}[t!]
\caption{Comparing the two variants of Alg.~\ref{alg:strat_sampling} regarding covering radius.}
\centering
\begin{tabularx}{\textwidth}{rRRRR}
\toprule
Dimension & w/ Lines 6--7 Wins & Ties & w/o Lines 6--7 Wins & Measure\\
\midrule
2 & 822 & 96  & 102 & exact $d_\mathrm{cr}$\\ 
3 & 610 & 96  & 314 & exact $d_\mathrm{cr}$\\ 
4 & 800 & 94  & 126 & exact $d_\mathrm{cr}$\\
5 & 698 & 120 & 202 & exact $d_\mathrm{cr}$\\
6 & 720 & 167 & 133 & upper bound\\
7 & 712 & 271 & 37 & upper bound\\
8 & 517 & 466 & 37 & upper bound\\
9 & 136 & 807 & 77 & upper bound\\
10 & 142 & 778 & 100 & upper bound\\ 
\bottomrule
\end{tabularx}
\label{tab:avoid_wins_losses}
\end{table*}

%


\appendix

\section*{Appendix}

Table~\ref{tab:avoid_wins_losses} compares the two variants of Alg.~\ref{alg:strat_sampling} regarding the covering radius. 
Point sets from 4 to 1024 points were tested in two to ten dimensions.
The variant containing lines 6--7 has an advantage in the majority of cases.

\bibliographystyle{spbasic}      
\bibliography{literature}   


\end{document}